\documentclass[12pt]{article}
\usepackage[dvips]{graphicx}

\def\nn{\nonumber}
\def\dfrac{\displaystyle\frac}
\def\numt#1#2{#1 \times 10^{#2}}
\def\etal{{\it et al.}}

\def\ie{{\it i.e.,~}}
\def\eg{{\it e.g.~}}
\def\bs{\bigskip}

\def\PR#1#2#3{Phys. Rev. {\bf #1}, #2 (#3)}
\def\PRL#1#2#3{Phys. Rev. Lett. {\bf #1}, #2 (#3)}
\def\PL#1#2#3{Phys. Lett. {\bf #1}, #2 (#3)}
\def\NP#1#2#3{Nucl. Phys. {\bf #1}, #2 (#3)}
\def\PTP#1#2#3{Prog. Theor. Phys. {\bf #1}, #2 (#3)}
\def\EPJ#1#2#3{Eur. Phys. J. {\bf #1}, #2 (#3)}

\def\eqref#1{eq.(\ref{eqn:#1})}
\def\Eqref#1{Equation(\ref{eqn:#1})}
\def\eqsref#1{eqs.(\ref{eqn:#1})}

\def\eqvref#1{(\ref{eqn:#1})}

\def\eqlab#1{\label{eqn:#1}}

\def\Fgref#1{Fig.\ref{fig:#1}}
\def\Figref#1{Figure\ref{fig:#1}}
\def\Fgsref#1{Figs.\ref{fig:#1}}

\def\Fgvref#1{\ref{fig:#1}}
\def\Fglab#1{\label{fig:#1}}

\def\l{\left}
\def\r{\right}
\def\gsim{~{\rlap{\lower 3.5pt\hbox{$\mathchar\sim$}}\raise 1pt\hbox{$>$}}\,}
\def\lsim{~{\rlap{\lower 3.5pt\hbox{$\mathchar\sim$}}\raise 1pt\hbox{$<$}}\,}
\makeatletter
\newtoks\@stequation

\def\subequations{\refstepcounter{equation}%
  \edef\@savedequation{\the\c@equation}%
  \@stequation=\expandafter{\theequation}%   %only want \theequation
  \edef\@savedtheequation{\the\@stequation}% %expanded once
  \edef\oldtheequation{\theequation}%
  \setcounter{equation}{0}%
  \def\theequation{\oldtheequation\alph{equation}}}

\def\endsubequations{%
  \ifnum\c@equation < 2 \@warning{Only \the\c@equation\space subequation
    used in equation \@savedequation}\fi
  \setcounter{equation}{\@savedequation}%
  \@stequation=\expandafter{\@savedtheequation}%
  \edef\theequation{\the\@stequation}%
  \global\@ignoretrue}

\def\eqnarray{\stepcounter{equation}\let\@currentlabel\theequation
\global\@eqnswtrue\m@th
\global\@eqcnt\z@\tabskip\@centering\let\\\@eqncr
$$\halign to\displaywidth\bgroup\@eqnsel\hskip\@centering
     $\displaystyle\tabskip\z@{##}$&\global\@eqcnt\@ne
      \hfil$\;{##}\;$\hfil
     &\global\@eqcnt\tw@ $\displaystyle\tabskip\z@{##}$\hfil
   \tabskip\@centering&\llap{##}\tabskip\z@\cr}

\makeatother
\def\cerenkov{$\check{\rm C}$erenkov~}
\def\sss{\scriptscriptstyle}
\def\atm{\sss{\rm ATM}}
\def\rct{\sss{\rm RCT}}
\def\sun{\sss{\rm SOL}}
\def\dmns{\delta_{\sss{\rm MNS}}}

\def\ssun#1{\sin^2 #1\theta_{\sss{\rm SOL}}}
\def\satm#1{\sin^2 #1\theta_{\sss{\rm ATM}}}
\def\srct#1{\sin^2 #1\theta_{\sss{\rm RCT}}}

\title{
Solving the degeneracy of
the lepton-flavor mixing angle
$\theta_{\atm}$
by the T2KK two detector neutrino oscillation experiment
}
\author{
Kaoru Hagiwara$^1$
and
Naotoshi Okamura$^2$\thanks{e-mail:~okamura@yukawa.kyoto-u.ac.jp} 
\\ \\
{\it \small
$^1$ KEK Theory Division and SOKENDAI,}
{\it \small Tsukuba, 305-0801 Japan } \\
{\it \small 
$^2$ Yukawa Institute for Theoretical Physics,
Kyoto University,}
{\it \small Kyoto, 606-8502 Japan}
}
\date{~}
\begin{document}
\maketitle
\vspace{-12.0cm}
\begin{flushright}
hep-ph/0611058\\
KEK-TH-1109\\
YITP-06-48
\end{flushright}
\vspace{ 12.0cm}
\vspace{-4.0cm}
\begin{abstract}
If the atmospheric neutrino oscillation amplitude,
$\satm{2}$ is not maximal,
there is a two fold ambiguity
in the neutrino parameter space:
$\satm{}>0.5$ or $\satm{}<0.5$.
 In this article, we study the impact of this degeneracy,
the so-called octant degeneracy, 
on the T2KK experiment, 
which is a proposed 
extension
of the T2K (Tokai-to-Kaimoka)
neutrino oscillation experiment with an additional
water \cerenkov detector placed in Korea.
We find that the degeneracy between 
$\satm{}=$ 0.40 and 0.60
can be resolved at the 3$\sigma$
level for 
$\srct{2}>0.12$ (0.08) for the optimal combination of
a $3.0^\circ$ off-axis beam (OAB) at SK ($L=295$km) and 
a $0.5^\circ$ OAB at $L=1000$km
with a far detector of 100kton volume,
after 5 years of exposure with $\numt{1.0~(5.0)}{21}$POT/year,
if the hierarchy is normal.
We also study
the influence of the octant degeneracy
on the capability of T2KK experiment
to determine the mass hierarchy and
the leptonic CP phase.
The capability of rejecting the wrong mass hierarchy grows with
increasing $\satm{}$ when the hierarchy is normal,
whereas it is rather insensitive to $\satm{}$ for the inverted
hierarchy.
We also find that the 1$\sigma$ allowed region of the CP phase
is not affected significantly even when the octant degeneracy
is not resolved.
All our results are obtained for the 22.5 kton Super-Kamiokande as
a near detector and without an anti-neutrino beam.
\end{abstract}

\section{Introduction}
\label{sec:1}
\setcounter{equation}{0}
A decade ago, it was difficult to believe that neutrinos have
mass and the lepton flavor mixing matrix, 
the Maki-Nakagawa-Sakata (MNS) matrix \cite{MNS},
has two large mixing angles
\cite{atm}-
\nocite{k2k, minos, CHOOZ, solar}
\cite{KamLAND}.
Within the three neutrino framework,
2 mass-squared differences,
3 mixing angles,
and
1 CP phase
can be resolved by neutrino oscillation experiments.
So far,
the magnitude of the larger mass-squared difference,
the magnitude and the sign of the smaller one,
two of the three mixing angles,
and the upper bound of the third mixing angle
have been known.
The sign of the larger mass-squared difference
(the mass hierarchy pattern),
the magnitude of the third mixing angle ($\theta_{\rct}$),
and the leptonic CP phase ($\dmns$)
are yet to be measured.

In the previous papers \cite{HOS1, HOS2},
we studied in detail the physics impacts of the idea \cite{HOSetc}
of placing a far detector in Korea along the T2K neutrino beam line.
For concreteness we examined the effects of placing 
a 100kton water \cerenkov detector in Korea,
about $L=1000$km away from J-PARC
(Japan Proton Accelerator Research Complex) \cite{j-parc},
during the T2K (Tokai-to-Kamioka) experiment period \cite{t2k},
which plans to accumulate 
$5\times10^{21}$ POT (protons on target) in 5 years.
We find that 
this experiment with two detectors for one beam,
which may be called the T2KK experiment \cite{korean},
can determine the mass hierarchy pattern,
by comparing the $\nu_{\mu} \to \nu_e$ transition probability 
measured at Super-Kamiokande (SK)
and that at a far detector in Korea.
Moreover,
both the sine and cosine of the CP phase can be measured
from the energy dependence of the $\nu_{\mu}^{} \to \nu_e^{}$
oscillation probability,
which can be measured by selecting the quasi-elastic
charged current events.
By studying these physics merits of the T2KK experiment 
semi-quantitatively,
we find an optimal combination of a $3^{\circ}$ off-axis beam (OAB)
at SK and a $0.5^{\circ}$ OAB in the east coast of Korea
at $L = 1000$km,
for which the mass hierarchy and the CP phase ($\dmns$)
can be detemined without invoking an 
anti-neutrino phase \cite{HOS1, HOS2},
when the mixing angle $\theta_{\rct}$ is not too small.
In the related study \cite{T2KK}, a grander
prospect of the T2KK idea has been explored,
where two identical huge detectors of several 100kton
volume is placed in Kamioka and in Korea,
and the future upgrade of the J-PARC beam intensity
has also been considered.
The idea of placing two detectors along one neutrino beam has also
been explored for the Fermi Lab. neutrino beam \cite{super-nova}.

 In this report, we focus on the yet another degeneracy
in the neutrino parameter space,
which shows up when the amplitude of
the atmospheric neutrino oscillation,
$\satm{2}$,
is not maximal.
 Hereafter, we call this degeneracy
between $\satm{}>0.5$ and $\satm{}<0.5$, or
between ``$90^\circ-\theta_{\atm}$'' and ``$\theta_{\atm}$'',
as the octant degeneracy \cite{octant}.

 Since the best fit value of the mixing angle
$\theta_{\atm}$ is $45^\circ$ \cite{atm,k2k,minos}, 
we set $\satm{2}=1$ in all our precious studies \cite{HOS1,HOS2},
and hence 
we did not pay attention on physics impacts of the octant degeneracy.
 However, we are concerned that the octant degeneracy affect the 
capability of the T2KK experiment for the mass hierarchy determination 
and the CP phase measurement,
 because the leading term of the $\nu_\mu^{}\to\nu_e^{}$ 
oscillation probability
is proportional to $\satm{}$, not $\satm{2}$.
 If the value of $\satm{2}$ is 0.99, which is 1\% smaller than the
maximal mixing,
 the value of $\satm{}$ is $\satm{}=0.45$ or $0.55$,
which differ by 20\%.

For $\satm{2}=0.92$, which is still allowed at the 90\% CL
\cite{atm,k2k,minos}, 
$\satm{}=0.64$ or 0.36, which differ by almost a factor of two.
 Therefore,
 we also examine impacts of varying $\satm{}$ 
on the mass hierarchy determination
and the CP phase measurement by T2KK.
In our semi-quantitatively analysis, we follow the strategy of 
ref.\cite{HOS1,HOS2} where we adopt SK as a near side detector
and postulate a 100 kton water \cerenkov detector at $L=1000$km,
and the J-PARC neutrino beam orientation is adjusted to 
$3.0^\circ$ at SK and $0.5^\circ$ at the Korean detector site. 

This article is organized as follows.
In section \ref{sec:of-ex},
we fix our notation and show how the octant degeneracy affects the
oscillation probabilities.
In section \ref{sec:chi},
we review our analysis method and
present an explicit form of the $\Delta \chi^2$ function
which we use to measure the capability of the T2KK experiment
semi-quantitatively.
In section \ref{sec:octant},
we show the results of our numerical calculation on the resolution
of the octant degeneracy.
In section \ref{sec:mass},
we examine the capability of the T2KK experiment
for the mass hierarchy determination,
in the presence of the octant degeneracy.
We also show the effect of the octant degeneracy
on the CP phase measurement in section \ref{sec:CP}.
In the last section,
we summarize our results and give discussions.
\section{Oscillation formular and experimental bound}
\label{sec:of-ex}
When a neutrino of flavor $\alpha$ is created at the neutrino source
with energy $E$,
it is a mixture of the mass eigenstates, $\nu_i^{}$
\begin{equation}
 \left|\nu_{\alpha} \right\rangle=
 \sum^{3}_{i=1} U_{\alpha i}~
 \left|\nu_{i} \right\rangle\,,
~\mbox{{\hspace*{3ex}}}(\alpha = e, \mu, \tau)
\end{equation}
where $U_{\alpha i}$ is the element of the Maki-Nakagawa-Sakata
(MNS) matrix \cite{MNS}.
Without loosing generality, we can take $U_{e2}$ and $U_{\mu 3}$
to be real and non-negative
and allow $U_{e3}$ to have a complex phase $\dmns$ \cite{HO1,PDB}.

After traveling the distance $L$ in the vacuum,
a neutrino flavor eigenstate
$\left| \nu_{\beta} \right\rangle$ is found 
with the probability
\begin{eqnarray}
 P_{\nu_\alpha \to \nu_\beta}
 &=& \left|
\left\langle \nu_\beta \right.
\left|\nu_{\alpha} (L)\right\rangle
\right|^2
=
\left| \sum_{j=1}^{3}
U_{\alpha j}^{}
\exp{\left(-i\dfrac{m_j^2}{2E}L\right)}
U_{\beta j}^{\ast}
\right|^2 \nn \\
&=&
\delta_{\alpha\beta}
-4\sum_{i>j}
 \Re(U^{\ast}_{\alpha i}U_{\beta i}^{}U_{\alpha j}^{}U^{\ast}_{\beta j})
 \sin^2\dfrac{\Delta_{ij}}{2} 
+2\sum_{i>j}
 \Im(U^{\ast}_{\alpha i}U_{\beta i}^{}U_{\alpha j}^{}U^{\ast}_{\beta j})
 \sin\Delta_{ij}\,,
\eqlab{prob1}
\end{eqnarray}
where $m_j$ is the mass of $\nu_i^{}$ 
and $\Delta_{ij}$ is
\begin{eqnarray}
\Delta_{ij} \equiv
 \dfrac{m_j^2 - m_i^2}{2E}L
\simeq
 2.534 \dfrac{\left(m_j^2 - m_i^2\right)[\mbox{{eV}}^2]}
{E[\mbox{{GeV}}]}
L\left[\mbox{km}\right]\,.
\eqlab{Delta}
\end{eqnarray} 
\Eqref{prob1} shows that neutrino flavor oscillation is governed by 
the two mass-squared differences
and the lepton number conserving combinations of the 
MNS matrix elements.

We take
$|\Delta_{13}| > |\Delta_{12}|$ without loosing generality.
Under this parameterization, 
atmospheric neutrino observation \cite{atm} and 
the accelerator based long baseline (LBL) experiments, 
K2K \cite{k2k} and MINOS \cite{minos},
which measure the $\nu_\mu^{}$ survival probability,
are sensitive to the magnitude of the larger mass-squared
difference and $U_{\mu 3}^{}$:
\begin{subequations}
\begin{eqnarray}
&\numt{1.5}{-3} {\mbox{{eV}}}^2 <
 |m_3^2 - m_1^2| <
 \numt{3.4}{-3} {\mbox{{eV}}}^2 \,,\\
\eqlab{datm-data}
&\satm{2} 
\equiv
4U_{\mu 3}^2 \left(1-U_{\mu 3}^2\right)
> 0.92\,,
\eqlab{satm-data}
\end{eqnarray}
\eqlab{atm-data}
\end{subequations}
$\!$each at the 90$\%$ confidence level.
Hereafter, we use $\satm{}$ instead of the $U_{\mu 3}^2$ for
brevity.

The reactor experiments,
which observe the survival probability of the $\bar{\nu}_e^{}$
at $L \sim 1$km from a reactor,
are sensitive to the value of the larger mass-squared difference
and the absolute value of $U_{e3}$.
The CHOOZ experiment \cite{CHOOZ}
reported no reduction of the $\bar{\nu}_e$ flux
and find 
 \begin{eqnarray}
 \srct{2}&\equiv&
 4\left|U_{e3}\right|^2 \left(1-\left|U_{e3}\right|^2\right)
< 
(0.20,~0.16,~0.14)\nn\\
&&\mbox{{~~~~~  for  ~}}
\l| m_3^2 - m_1^2 \r| = \numt{(2.0,~2.5,~3.0)}{-3}\mbox{{eV}}^2\,,
\eqlab{rct-data}
\end{eqnarray}
$\!\!$at the 90\% confidence level.
In the following, we denote $|U_{e3}|^2$ as 
$\srct{}$.

 The solar neutrino observations \cite{solar}, 
and
 the KamLAND experiment \cite{KamLAND},
which measure the survival probability of 
$\nu_e^{}$ and $\bar{\nu}_e^{}$, respectively,
are sensitive to the smaller mass-squared difference
and the value of $U_{e2}$.
The present constraints can be expressed as
\begin{subequations}
\begin{eqnarray}
&m^2_2 - m^2_1 =  (8.0 \pm 0.3) \times 10^{-5} {\mbox{{eV}}}^2 \,,\\
&\ssun{} = 0.30 \pm 0.03\,. 
\end{eqnarray}
\eqlab{sun-data}
\end{subequations}
$\!\!$The sign of $m^2_2 - m^2_1$ is determined by
the matter effect in the sun \cite{mat-eff, msw}.
In these experiments, the order of $\Delta_{12}$ is roughly 1 and
the terms with $\Delta_{13}$ oscillate quickly within the experimental
resolution of $L/E$.
After averaging out the contribution from $\Delta_{13}$, and
neglecting terms of order $\srct{}$, 
we obtain the relation;
\begin{equation}
\ssun{2} = 
 4U_{e1}^{2}U_{e2}^{2} = 
 4U_{e2}^{2}(1-U^2_{e2}-\l|U_{e3}^2\r|)\,.
\eqlab{sun-rel}
\end{equation}
These simple identification,
\eqsref{satm-data}, \eqvref{rct-data}, and \eqvref{sun-rel},
are found to give a reasonably good description of the present data
in dedicated studies \cite{lisi}
of the experimental constraints in the three neutrino
model.
In this paper, we parameterize the CP phase as \cite{PDB}
\begin{equation}
\dmns = - \arg U_{e3} \,.
\end{equation}
The other elements of the MNS matrix can be obtained form the 
unitary conditions \cite{HO1}.
This convention allows us to express the MNS matrix directly
in terms of the three observed amplitudes.

 The probability of the neutrino oscillation, \eqref{prob1}, is
modified by the matter effect \cite{mat-eff, msw}, 
 because only $\nu_e^{}$ and $\bar{\nu}_e^{}$ feel the potential
by the extra charged current interactions with the electron inside the
matter.
 This extra potential for $\nu_e^{}$ is written as
\begin{equation}
 a = 2\sqrt{2} G_F E_\nu^{} n_e
  \simeq \numt{7.56}{-5} \mbox{{[eV$^2$]}} 
  \left(\dfrac{\rho}{\mbox{{g/cm$^3$}}}\right)
  \left(\dfrac{E_\nu}{\mbox{{GeV}}}\right)\,,
\eqlab{matt_a}
\end{equation}
where 
$G_F$ is the Fermi coupling constant, 
$E_\nu$ is the neutrino energy,
$n_e$ is the electron number density,
and
$\rho$ is the matter density.
The extra potential for $\bar{\nu}_e$ has the opposite sign.
 Because the matter effect is small at low energies
and also because the phase factor $\Delta_{12}$ is small near the
first oscillation maximum, $\Delta_{13}\sim\pi$,
we find that an approximation of keeping terms linear in the
matter effect and $\Delta_{12}$
is useful for analyzing the LBL experiments at sub GeV to
a few GeV region \cite{HOS1, HOS2, KOST05,HKOT1}:
\begin{subequations}
\begin{eqnarray}
P_{\nu_{\mu}^{} \rightarrow \nu_e^{}}
 &=&
 2(1+q)\srct{} \left( 1 + A^{e} \right)
\sin^2 \left( \dfrac{\Delta_{13}}{2} + B^e \right) \,,
\eqlab{prob_me}\\
P_{\nu_{\mu}^{} \rightarrow \nu_{\mu}^{}}
 &=&
1 - (1-q^2)\left( 1 + A^{\mu} \right)
\sin^2 \left( \dfrac{\Delta_{13}}{2} + B^{\mu} \right)\,,
\eqlab{prob_mm}
\end{eqnarray}
\eqlab{prob-m}
\end{subequations}
$\!\!$where 
\begin{equation}
\satm{} = \dfrac{1+q}{2}\,,
\eqlab{def_q}
\end{equation}
and
$A^{\alpha}$ and $B^{\alpha}$ are the correction
terms to the amplitude and the oscillation phase, respectively.
For $\alpha = e$, we find 
\begin{subequations}
\begin{eqnarray}
A^e 
&=& \dfrac{aL}{\Delta_{13}E} \cos 2\theta_{\rct} 
 -\dfrac{\Delta_{12}}{2}
  \dfrac{\sin2\theta_{\sun}}{\sin\theta_{\rct}}
  \sqrt{\dfrac{1-q}{1+q}}
  \sin \dmns\,, \eqlab{Ae}\\
B^e
&=& - \dfrac{aL}{4E}\cos 2\theta_{\rct}
 + \dfrac{\Delta_{12}}{2}
 \left(
  \dfrac{\sin2\theta_{\sun}}{2\sin\theta_{\rct}}
  \sqrt{\dfrac{1-q}{1+q}}
   \cos \dmns - \ssun{} \right) \eqlab{Be} \,.
\end{eqnarray}
\eqlab{ABe}
\end{subequations}
$\!\!$The octant degeneracy between ``$\theta_{\atm}$'' and
``$90^\circ-\theta_{\atm}$''
corresponds to the degeneracy in the sign of $q$.
When $q$ denotes the true value for the octant degeneracy,
$-q$ is its fake value.

 Using typical numbers of the parameters from the atmospheric
neutrino observation and LBL experiments, \eqref{atm-data},
and those from the solar neutrino observation and
the KamLAND experiment, \eqref{sun-data},
the $\nu_\mu \to \nu_e$ transition probability
can be expressed as
\begin{subequations}
\begin{eqnarray}
 P_{\nu_{\mu}^{} \rightarrow \nu_e^{}}
 &\sim&
 0.05 \left( 1+q \right)\left(\dfrac{\srct{2}}{0.10}\right)
 \left( 1 + A^{e} \right)
\sin^2 \left(\dfrac{\Delta_{13}}{2} + B^e \right) \,.
\eqlab{prob_me2}\\
A^{e} &\sim& 0.37
\left(\dfrac{\pi}{\Delta_{13}}\right)
\left(\dfrac{L}{1000{\rm [km]}}\right)
-\left[ 0.29
\sqrt{\dfrac{1-q}{1+q}}
\left( \dfrac{0.10}{\srct{2}} \right)^{1/2}
\sin\dmns
\right]
\dfrac{|\Delta_{13}|}{\pi} \,,
\eqlab{Ae2}
\\
B^{e} &\sim&
- 0.29\left( \dfrac{L}{1000{\rm [km]}} \right)
+\left[0.15
\sqrt{\dfrac{1-q}{1+q}}
\left( \dfrac{0.10}{\srct{2}} \right)^{1/2}
\cos \dmns
- 0.015 \right] 
\dfrac{|\Delta_{13}|}{\pi}
\eqlab{Be2}\,,~~~~~~~
\end{eqnarray}
\eqlab{ABe2}
\end{subequations}
$\!\!$around the oscillation maximum, $|\Delta_{13}|\sim\pi$.
Since the amplitude is proportional to $\satm{}=(1+q)/2$,
we expect that
the octant degeneracy can be solved by
measuring the $\nu_\mu \to \nu_e$ transition probability, 
{\em if} the value of the $\srct{2}$ is known precisely.
Because the first term of $A^e$
changes sign according to the mass hierarchy pattern,
$\Delta_{13}\sim \pi$ for the normal and
$\Delta_{13} \sim -\pi$ for the inverted,
the amplitude of the transition probability is sensitive
to the mass hierarchy pattern.
The difference between the two hierarchy cases grows with 
the baseline length when $L/E$ is fixed at 
around the oscillation maximum 
\cite{HOS1,HOS2}.
If there is only one detector at $L\sim O(100)$km,
the small difference from the matter effect
can be absorbed by the sign of $q$ in the leading term of
\eqref{prob_me2}.

The $q$-dependence in $A^e$ and $B^e$
in \eqsref{Ae2} and \eqvref{Be2}
may seem to affect the measurement of
the leptonic CP phase.
We find, however, that the $q$-dependence of the
coefficient of the CP phase in the $\nu_\mu \to \nu_e$
transition probability is not strong, because
\begin{equation}
 \left( 1+q \right) \sqrt{\dfrac{1-q}{1+q}}
=
\sqrt{1-q^2}\,,
\eqlab{q-CP}
\end{equation}
which is independent of the octant degeneracy.
Our numerical studies presented below confirms the validity
of the above approximations.

Around the first dip of the $\nu_\mu$ survival probability
$|\Delta_{13}|\sim\pi$,
we find
\begin{subequations}
\begin{eqnarray}
A^{\mu} &\sim&
0.018
\left(\dfrac{q}{1-q}\right)
\left(\dfrac{\pi}{\Delta_{13}}\right)
\left(\dfrac{L}{1000{\rm [km]}}\right)
\left( \dfrac{\srct{2}}{0.10} \right)
\eqlab{Am2}\,, \\
B^{\mu} &\sim&
0.014
\left(\dfrac{q}{1-q}\right)
\left(\dfrac{L}{1000{\rm [km]}}\right)
\left( \dfrac{\srct{2}}{0.10} \right)
\nn \eqlab{Bm2} \\
&&\hspace*{15ex}-
\left[
0.037 - 0.008
\left( \dfrac{\srct{2}}{0.10} \right)^{1/2}
\cos \dmns
\right]
\dfrac{|\Delta_{13}|}{\pi}
\,.
\end{eqnarray}
\eqlab{ABm2}
\end{subequations}
$\!\!\!$Although the shift in the amplitude $A^\nu$ and
that in the phase $B^\mu$ are both proportional to $q$, 
their magnitudes are found to be less than 0.7\% and 0.5\%,
respectively,
for $|q|<0.28$, \eqref{satm-data}.
Our numerical results confirm that the measurement of the 
$\nu_\mu \to \nu_\mu$ survival probability does not contribute
significantly to the resolution of the octant degeneracy.
On the other hand the smallness of the deviation from the leading
contribution
allows us to constrain $|m_3^2-m_1^2|$ and $\satm{2}$ accurately
by measuring the $\nu_\mu \to \nu_\mu$ survival probability.

\section{Analysis method}
\label{sec:chi}
 In this section, we explain how we treat signals and backgrounds
in our numerical analysis, and introduce a $\chi^2$ function 
which measures the capability of the T2KK experiment
semi-quantitatively.
We consider a water \cerenkov detector at Korea in this study,
because it allows us to distinguish 
clearly the $e^{\pm}$ events from $\mu^{\pm}$ events.
The fiducial volume of the detector placed at Korea 
is assumed 100~kton, which is roughly 5
times larger than that of SK, 22.5 kton,
in order to compensate for the longer base-line length.
We use only the CCQE events in our analysis,
because they allow us to reconstruct the neutrino energy event by
event \cite{k2k}.
Since the Fermi-motion of the target nucleon would dominate 
the uncertainty of the neutrino energy reconstruction,
which is about 80 MeV \cite{k2k},
we take the width of the energy bin as 
$\delta E_\nu=200$~MeV for $E_\nu > 400$ MeV.
 The signals in the $i$-th energy bin,
$E^i_{\nu}
\equiv (200{\rm MeV} \times i) < E^{}_{\nu}
<E^i_{\nu}+\delta E_\nu^{}$,
are then calculated as
\begin{equation}
N_\alpha^{i} (\nu_\mu)=
 M N_A
 \int_{E_\nu^i}^{E_\nu^{i}+\delta E_\nu^{}}
 \Phi_{\nu_\mu}(E)~
 P_{\nu_\mu\to\nu_\alpha}(E)~ 
 \sigma_\alpha^{QE}(E)~
 dE\,,
\eqlab{N}
\end{equation}
where
$P_{\nu_{\mu}\rightarrow \nu_{\alpha}}$
is the neutrino oscillation probability including the matter effect, 
$M$ is the detector mass,
$N_{A} = 6.017\times10^{23}$ is the Avogadro constant,
$\Phi_{\nu_\mu}$ is the $\nu_{\mu}$ flux from J-PARC \cite{ichikawa},
and
$\sigma_\alpha^{QE}$ is the CCQE cross section
per nucleon in water \cite{k2k}.
For simplicity, the detection efficiencies of both detectors for both
$\nu_{\mu}$ and $\nu_e$ CCQE events are set at $100\%$.

We consider the following background events 
for the signal of $e$-like events ($\alpha=e$)
and $\mu$-like events ($\alpha=\mu$), 
\begin{eqnarray}
N_{\alpha}^{i,{\rm BG}}
&=&
N_{\alpha}^{i}(\nu_e) +
N_{\bar{\alpha}}^{i}(\bar{\nu}_e) +
N_{\bar{\alpha}}^{i}(\bar{\nu}_\mu)\,,
{\mbox{\hspace*{5ex}}}
(\alpha = e\,, \mu)\,,
\eqlab{BG_e}
\end{eqnarray}
respectively.
 The three terms correspond to the contribution from the secondary neutrino 
flux of the  $\nu_\mu$ primary beam,
which are calculated as in \eqref{N}
where $\Phi_{\nu_{\mu}}(E)~$ is replaced by $\Phi_{\nu_{\beta}}(E)$ for
$\nu_{\beta} = \nu_e\,,\bar{\nu}_e\,,\bar{\nu}_{\mu}$.
All the primary as well as secondary fluxes used in our analysis are
obtained from the web-site \cite{ichikawa}.
After summing up these background events,
the $e$-like and $\mu$-like events for the $i$-th bin are obtained as
\begin{equation}
\label{signal}
 N_{\alpha}^{i} = N_{\alpha}^{i}(\nu_\mu) + N_{\alpha}^{i,{\rm BG}}
\,,
{\mbox{\hspace*{5ex}}}
(\alpha = e\,, \mu)\,,
\end{equation}
respectively.

Our interest is the potential of the T2KK experiment for
solving the octant degeneracy and its influence on the 
resolution of the other degeneracies.
In order to quantify its capability,
we introduce a $\chi^2$ function,
\begin{equation}
\Delta\chi^2 \equiv
 \chi^2_{\rm SK}
+ \chi^2_{\rm Kr}
+ \chi^2_{\rm sys} 
+ \chi^2_{\rm para}\,,
\eqlab{def_chi^2}
\end{equation}
which measures the sensitivity of the experiment on
the model parameters.
The first two terms, $\chi^2_{\rm SK}$ and $\chi^2_{\rm Kr}$,
measure the parameter dependence of the fit to the SK and the Korean 
detector data, respectively,
\begin{eqnarray}
\eqlab{stat_chi2}
 \chi^2_{\rm SK,Kr}
 = \sum_{i} \left\{
\left(
\displaystyle\frac
{(N_e^{i})^{\rm fit} - (N_e^{i})^{\rm input}}
{\sqrt{(N^i_e)^{\rm input}}}
\right)^2
+
\left(
\displaystyle\frac
{(N_\mu^{i})^{\rm fit} - (N_\mu^{i})^{\rm input}}
{\sqrt{(N^i_{\mu})^{\rm input}}}
\right)^2
\right\}\,,
\end{eqnarray}
where $N^i_{\mu,e}$ is the calculated number of events
in the $i$-th bin,
and its square root gives the statistical error.
Here the summation is over all bins
from 0.4 GeV to 5.0 GeV for $N_{\mu}$,
0.4 GeV to 1.2 GeV for $N_{e}~$at SK,
and 0.4 GeV to 2.8GeV for $N_{e}~$at Korea.
In this energy region, we can include the second peak 
contribution in our analysis at Korea.
We include the contribution of the $\mu$-like events in order to constrain
the absolute value of $\Delta_{13}$  strongly in this analysis,
because a small error of $\Delta_{13}$ dilutes the phase shift $B^e$
\cite{HOS1, HOS2, KOST05}.

$N_{i}^{\rm fit}$ is calculated by allowing the model parameters
to vary freely and by including the systematic errors.
We take into account four types of the systematic errors in this
analysis.
The first systematic error is for the uncertainty in the matter density,
for which we allow $3\%$ overall uncertainty along the baseline, 
independently for T2K ($f^{\rm SK}_{\rho}$) and
the Tokai-to-Korea experiment ($f^{\rm Kr}_{\rho}$):
\begin{eqnarray}
\rho_{i}^{\rm fit} &=& f^{D}_{\rho}\,\rho^{\rm input}_i 
\hspace{5ex}
(D = {\rm SK,~Kr}) \,.
\eqlab{sys_matt}
\end{eqnarray}
The second ones are for the overall normalization of each neutrino flux,
for which we assume $3\%$ errors,
\begin{eqnarray}
 f_{\nu_\beta} &=& 1 \pm 0.03\,,  
 \eqlab{sys_flux}
\end{eqnarray}
for ($\nu_{\beta} = \nu_{e},~\bar{\nu}_{e},\nu_{\mu},~\bar{\nu}_{\mu}$),
which are taken common for T2K and the Tokai-to-Korea experiment.
The third ones are for the CCQE cross sections,  
\begin{equation}
\l(\sigma^{\rm QE}_{\alpha^{}}\r)^{\rm fit} = f^{\rm QE}_{\alpha}\, 
\l(\sigma^{\rm QE}_{\alpha^{}}\r)^{\rm input}\,,
\end{equation}
where $\alpha^{}$ denotes 
$\ell \equiv e=\mu$ and $\bar{\ell} \equiv \bar{e}=\bar{\mu}$.
Because $\nu_e$ and $\nu_\mu$ CCQE cross sections are expected to be very 
similar theoretically, we assign a common overall error of $3\%$ for 
$\nu_e$ and $\nu_{\mu}$ 
and an independent $3\%$ error for $\bar{\nu}_e$ and $\bar{\nu}_\mu$ CCQE 
cross sections.
The last one is the uncertainty of the fiducial volume,
for which we assign $3\%$ error independently for T2K 
($f_{\rm V}^{\rm SK}$) and the
Tokai-to-Korea experiment ($f_{\rm V}^{\rm Kr}$).
$N_{\alpha}^{i,{\rm fit}}$ is then calculated as
\begin{eqnarray}
\left[ N_\alpha^{i,{\rm fit}}(\nu_\beta) \right]_{\rm at~SK, Kr}&=&
f_{\nu_\beta^{}}\, 
f_{\alpha}^{\rm QE}\, 
f_{\rm V}^{\rm SK,Kr}\,
N_\alpha^{i}(\nu_\beta)\,, 
\end{eqnarray}
and accordingly, $\chi^2_{\rm sys}$ has four terms;
\begin{equation}
\eqlab{sys_chi2}
 \chi^2_{\rm sys} = 
\sum_{\beta = e,\bar{e},\mu,\bar{\mu}}
\left(
\dfrac{f_{\nu_{\beta}}-1}{0.03}
\right)^2
+
\sum_{\alpha = \ell, \bar{\ell}}
\left(
\dfrac{f^{\rm CCQE}_{\alpha}-1}{0.03}
\right)^2
+
\sum_{ D = {\rm SK,~Kr}}
\left\{
\left(
\dfrac{f^{D}_{\rho}-1}{0.03}
\right)^2
+
\left(
\dfrac{f^{D}_{\rm V}-1}{0.03}
\right)^2
\right\}\,.
\end{equation}
To put them shortly,
we account for 4 types of uncertainties
which are all assigned $3\%$ errors:
the effective matter density along each base line,
the normalization of each neutrino flux,
the CCQE cross sections for $\nu_l$ and $\bar{\nu}_l$,
and for the fiducial volume of SK, and that of the Korean detector.
In total, our $\Delta\chi^2$ function depends on 16 parameters,
the 6 model parameters and the 10 normalization factors.

Finally, $\chi^2_{\rm para}$ accounts for external constraints 
on the model parameters:
\begin{eqnarray}
\eqlab{para_chi2}
\chi^2_{\rm para}
&=&
\l(
\dfrac{
\l(m_2^2 - m_1^2\r)^{\rm fit} -
\numt{8.2}{-5}\mbox{{eV}$^2$}}
{ 0.6 \times 10^{-5}}
\r)^2
+
\l(\dfrac{\ssun{2}^{\rm fit}- %\ssun{2}^{\rm input}
0.83}{0.07}\r)^2 
\nn\\
&&+
\l(\dfrac{\srct{2}^{\rm fit}- \srct{2}^{\rm input}}{0.01}\r)^2\,.
\eqlab{chi-para}
\end{eqnarray}
The first two terms correspond to the present experimental constraints
from solar neutrino oscillation and KamLAND summarized in
\eqref{sun-data} \footnote{{%
The most recent results, \eqref{sun-data},
are slightly different from our inputs.
Because our analysis is not sensitive to the difference,
we use these values for the sake of keeping the consistency
with our previous studies \cite{HOS1,HOS2}.
}}.
In the last term, 
we assume that the planned future reactor experiments
\cite{KASKA}
should measure $\srct{2}$ with the expected uncertainty of 0.01.

\section{Octant degeneracy and the T2KK experiment}
\label{sec:octant}
In this section, we show the potential of the T2KK experiment for solving
the octant degeneracy and 
investigate the role of the far detector and
the future reactor experiments.
We show in \Fgref{oct.chi.wirct} the minimum $\Delta \chi^2$
as a function of $\satm{}$ expected at the T2KK experiment
after 5 years of data taking ($\numt{5}{21}$ POT).
The event numbers are calculated for a combination of 
$3.0^\circ$ OAB at SK and $0.5^\circ$ OAB at $L=1000$km
for the following parameters :
\begin{subequations}
\begin{eqnarray}
(m^2_3 - m^2_1)^{\rm input} &=&
 \numt{2.5}{-3} \mbox{{eV$^2$ (normal hierarchy)}}\,,
\eqlab{input_dm13}\\
(m^2_2 - m^2_1)^{\rm input} &=&
 \numt{8.2}{-5} \mbox{{eV$^2$}}\,, 
\eqlab{input_dm12}\\
\srct{2}^{\rm input} &=& 0.10\,, \eqlab{input_srct} \\
\ssun{2}^{\rm input} &=& 0.83\,, \eqlab{input_ssun} \\
\dmns^{\rm input} &=& 0^\circ\,, \pm 90^\circ\,,180^\circ\,,
\eqlab{input_dmns}\\
\rho^{\rm input}&=&3.0 \mbox{{g/cm$^3$ for $L=1000$km}}\,,
\eqlab{input_mden_Kr} \\
\rho^{\rm input}&=&2.8 \mbox{{g/cm$^3$ for SK}}\,.
\eqlab{input_mden_SK} 
\end{eqnarray}
\eqlab{input_set}
\end{subequations}
$\!\!$In the left-hand figure of \Fgref{oct.chi.wirct},
we show the cases for the input values
$\satm{}^{\rm input}=$0.35 (a), 0.40 (b), 0.45 (c) 
and 
in the right-hand figure, 
for $\satm{}^{\rm input}=$0.55 (d), 0.60 (e), 0.65 (f).
In each cases,
the fit has been performed by surveying the whole parameter space.
\begin{figure}[tb]
\centering
  \includegraphics[scale=0.6]{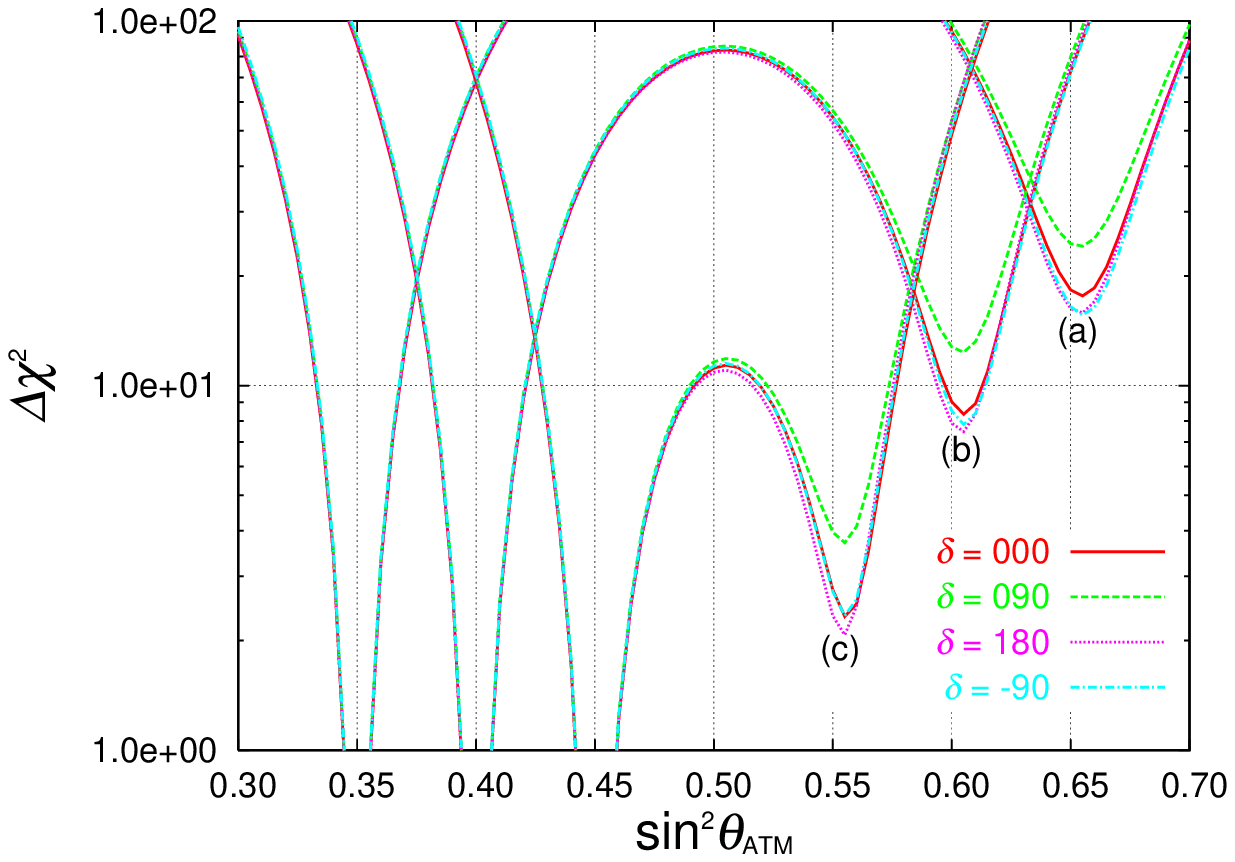}
~~~
  \includegraphics[scale=0.6]{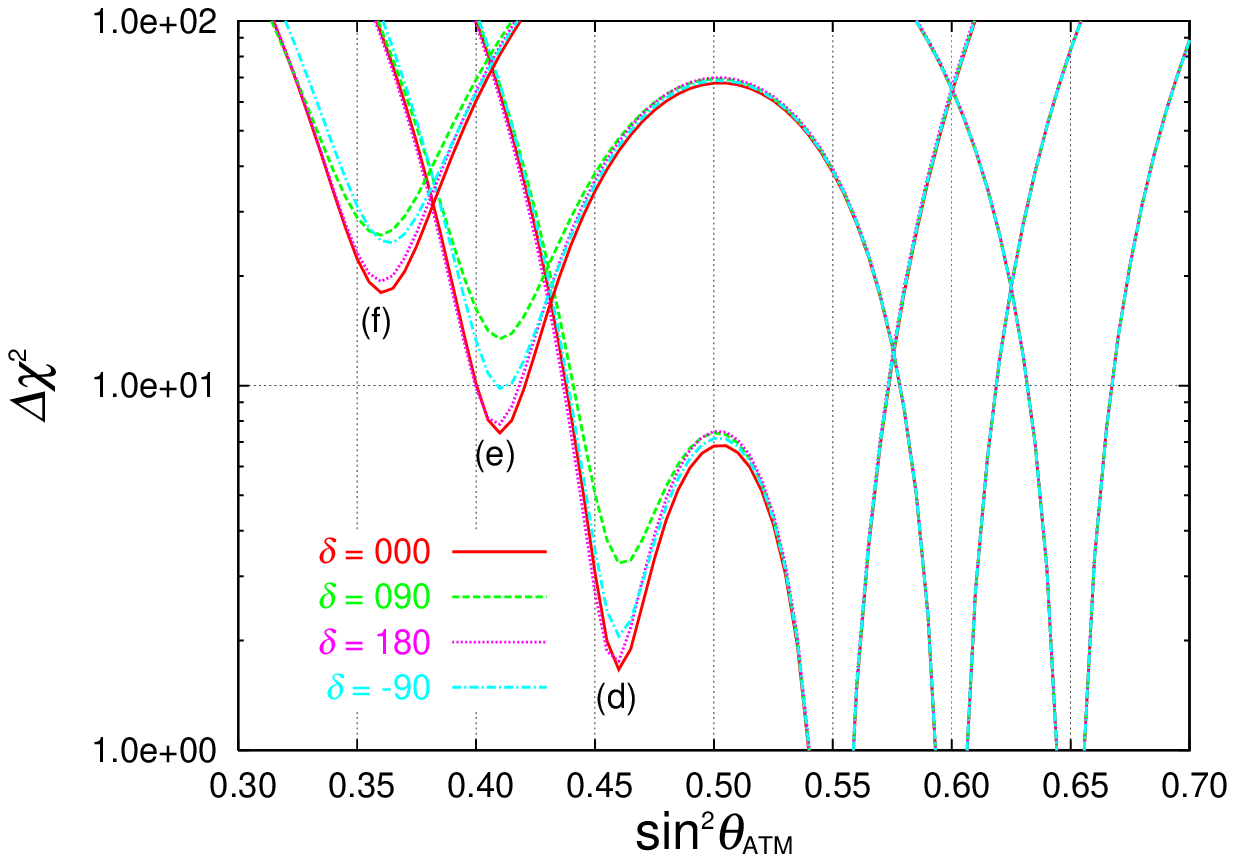}
 \caption{{%
 Minimum $\Delta \chi^2$ of the T2KK experiment as a function of
$\satm{}$.
The event numbers are calculated for a combination of 
$3.0^\circ$ OAB at SK and $0.5^\circ$ OAB at $L=1000$km
with 100 kton water \cerenkov detector, after 5 years running
($\numt{5}{21}$ POT).
The input parameters are chosen as in \eqref{input_set}.
In the left-hand figure,
$\satm{}^{\rm input}=0.35$ (a), $0.40$ (b), $0.45$ (c)
and 
in the right-hand figure,
$\satm{}^{\rm input}=0.55$ (d), $0.60$ (e), $0.65$ (f).
}}
\Fglab{oct.chi.wirct}
\end{figure}
We find from \Fgref{oct.chi.wirct},
that the octant degeneracy can be solved by T2KK experiment
when $\satm{2}=0.91$ \ie between $\satm{}=0.35$ and 0.65
at $4\sigma$.
For  $\satm{2}=0.96$ the degeneracy between $\satm{}=0.40$ and 0.60
can be resolved with $\Delta \chi^2 \geq 7$, or at 2.6$\sigma$.
However, it is difficult to solve the octant degeneracy
for $\satm{2}=0.99$, between $\satm{}=0.45$ and 0.55.

In the left-hand figure of \Fgref{oct.chi.wirct}, 
the minimum $\Delta \chi^2$ for $\delta=90^\circ$
is larger than those for the other CP phases.
In the right-hand figure,
the minimum $\Delta \chi^2$ is also largest at $\delta=90^\circ$.
There, however, the minimum $\Delta \chi^2$ for $\delta = -90^\circ$
is slightly larger than those for $\delta =0^\circ,180^\circ$.

\begin{figure}[t]
\centering
  \includegraphics[scale=0.6]{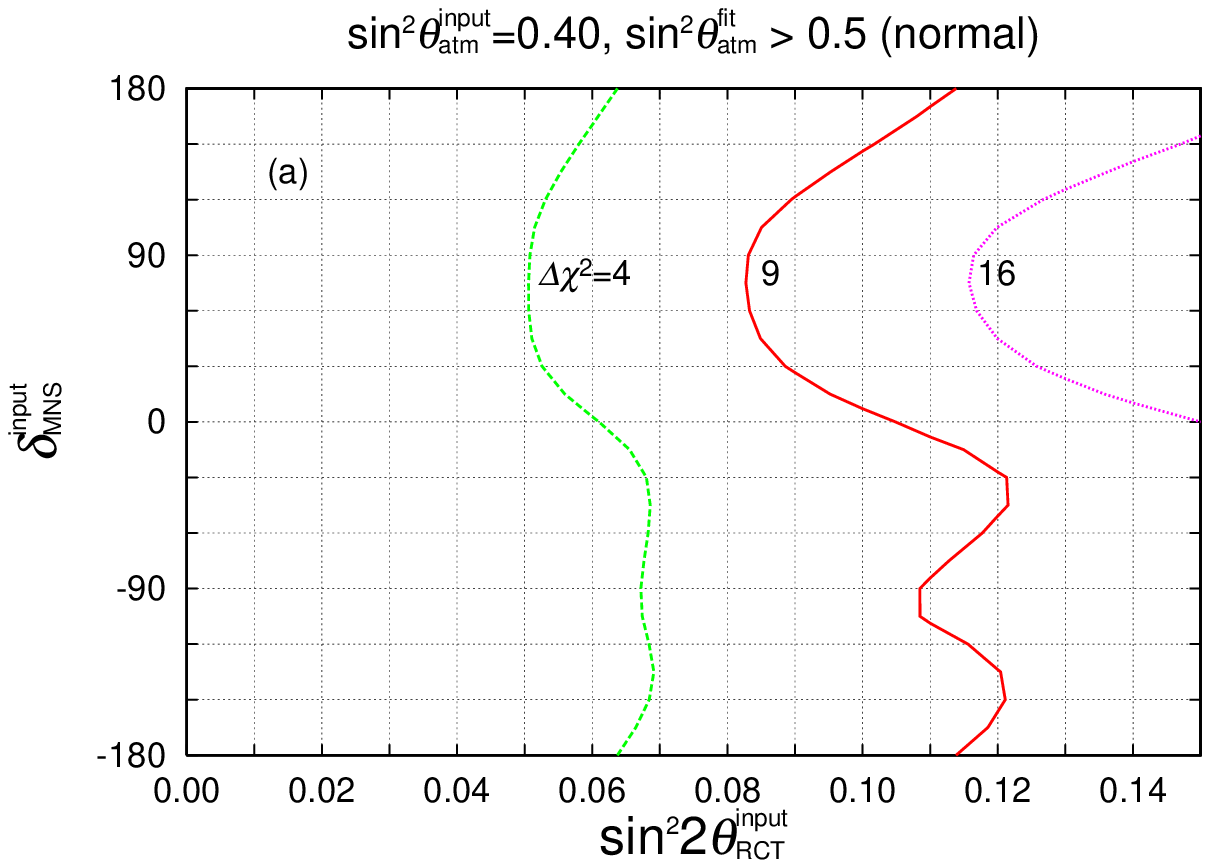}
~~~
  \includegraphics[scale=0.6]{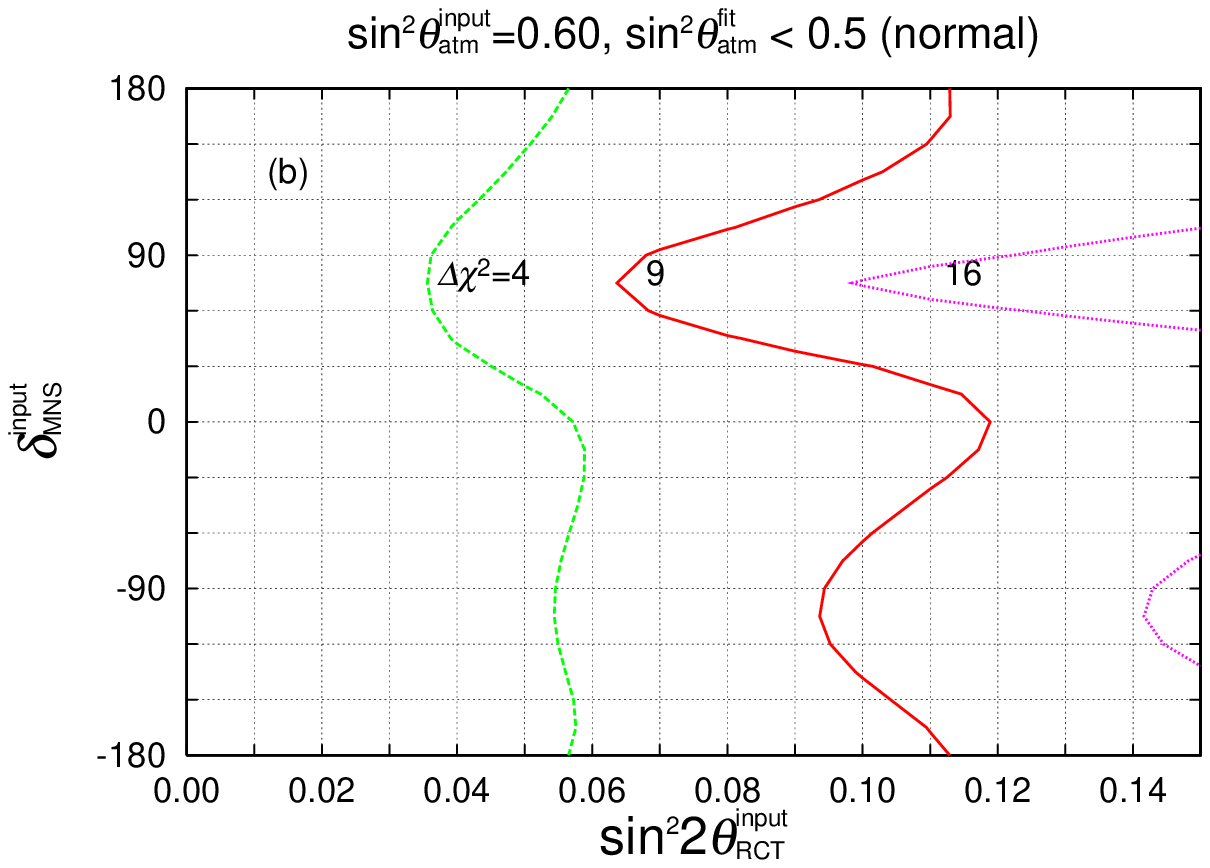}

  \includegraphics[scale=0.6]{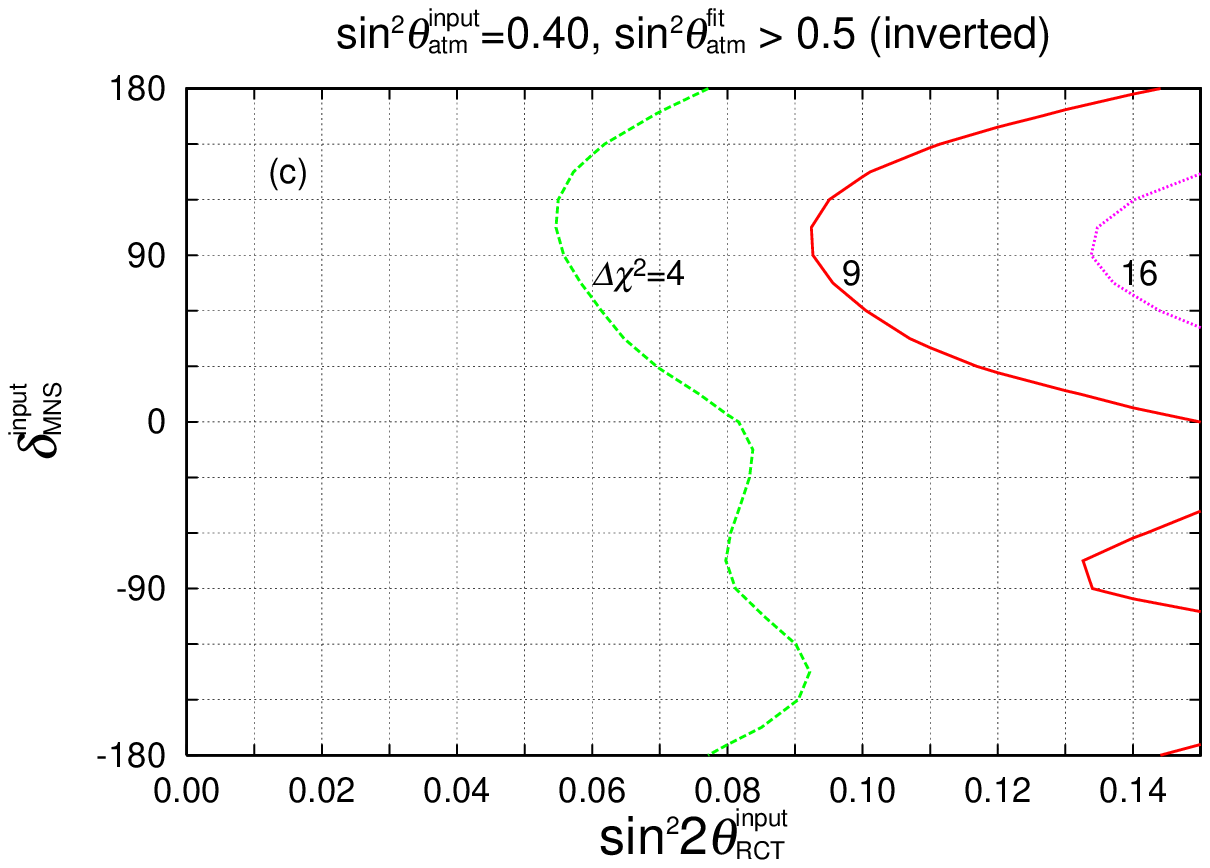}
~~~
  \includegraphics[scale=0.6]{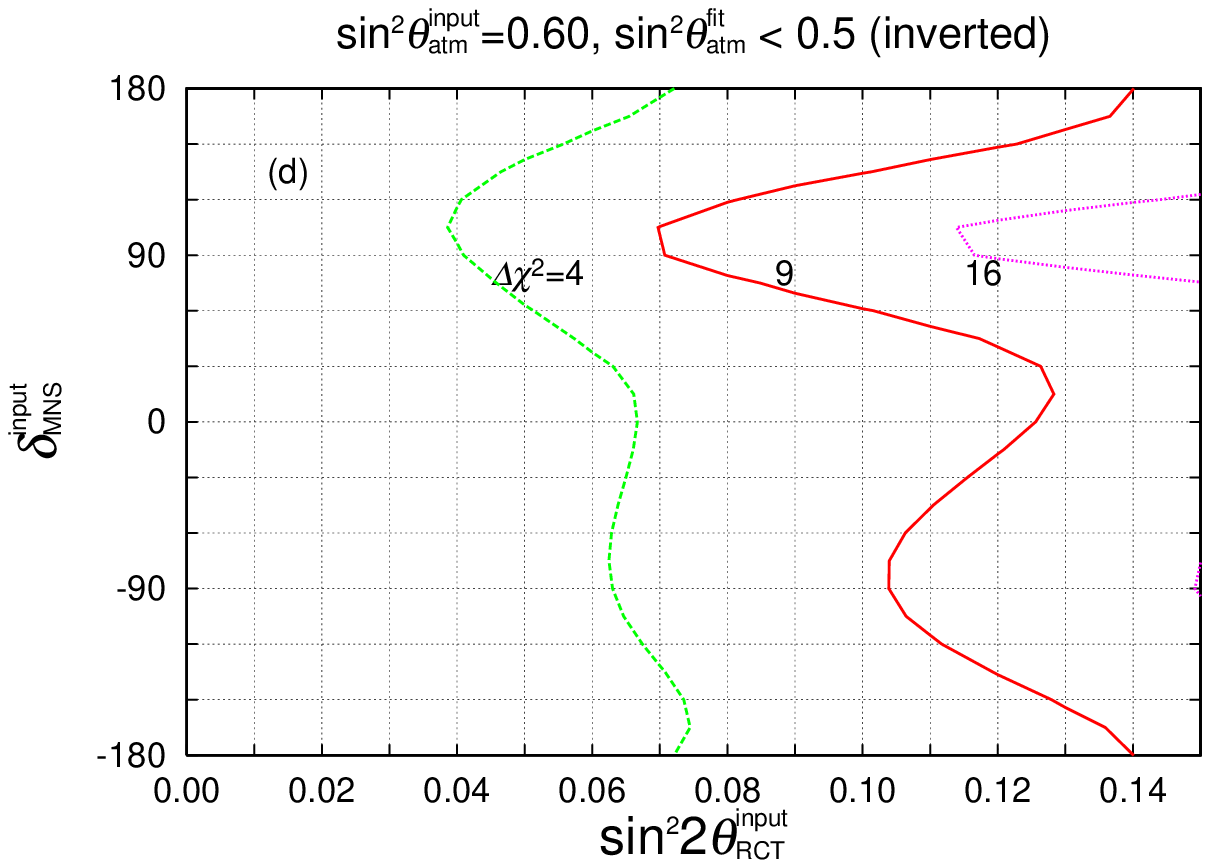}
 \caption{{%
 The potential of the T2KK experiment to solve the octant degeneracy
with the same OAB combination of the \Fgref{oct.chi.wirct}.
In each figures,
the event numbers are obtained for the model
parameters at various $\srct{2}^{\rm input}$,
$\dmns^{\rm input}$ and
$\satm{}^{\rm input}=0.40$ for (a,c), 
or
$\satm{}^{\rm input}=0.60$ for (b,d), 
with the normal (a,b) or the inverted (c,d) hierarchy. 
The other input parameters are same as those of \Fgref{oct.chi.wirct}.
All the parameters are taken freely in the fit 
under the constraint 
$\satm{}^{\rm fit}>0.50$ (a,c), 
or
$\satm{}^{\rm fit}<0.50$ (b,d).
The resulting values of minimum $\Delta \chi^2$ are shown as contours
for $\Delta \chi^2=$4, 9, 16.
}
}\Fglab{oct.cont}
\end{figure}
In order to explore the $\dmns$ dependence of
the capability of the T2KK experiment
to solve the octant degeneracy,
we show in \Fgref{oct.cont} contours of the
minimum $\Delta \chi^2$ in the whole space of
$\srct{2}^{\rm input}$ and $\dmns^{\rm input}$.
The event numbers are calculated
for various $\srct{2}^{\rm input}$ and $\dmns^{\rm input}$
values in each figure, with 
$\satm{}^{\rm input}=0.40$ for (a) and (c),
or
$\satm{}^{\rm input}=0.60$ for (b) and (d).
The other model parameters are set as in \eqref{input_set}.
\Fgsref{oct.cont} (a) and (b) are for the normal hierarchy,
$m_3^2-m_1^2=\numt{2.5}{-3}$eV$^2$, and
\Fgsref{oct.cont} (c) and (d) are for the inverted hierarchy,
$m_3^2-m_1^2=-\numt{2.5}{-3}$eV$^2$.
In performing the fit, all the 16 parameters
(6 model parameters and 10 normalization factors)
are varied freely
under the following constraints:
$\satm{}^{\rm fit}>0.5$ for (a) and (c),
$\satm{}^{\rm fit}<0.5$ for (b) and (d),
$(m_3^2-m_1^2)^{\rm fit} > 0$ for (a) and (b),
$(m_3^2-m_1^2)^{\rm fit} < 0$ for (c) and (d).
From \Fgsref{oct.cont}(a) and \Fgvref{oct.cont}(c),
we find that
$\satm{}^{\rm input}=0.40$ can be distinguished from
$\satm{}^{\rm fit} > 0.5$ at
$\Delta \chi^2 > 9$ (4) for
$\srct{2}^{\rm input}\gsim0.12$ (0.09)
when the normal (inverted) hierarchy is realized.
\Fgsref{oct.cont}(b) and \Fgvref{oct.cont}(d) show
that the octant degeneracy can be solved at 
$\Delta \chi^2 > 9$ for $\srct{2}^{\rm input}\gsim0.12$ (0.14) 
when $\satm{}^{\rm input}=0.60$ for the normal (inverted) hierarchy.

It is found in \Fgsref{oct.cont}(a) and \Fgvref{oct.cont}(b) that
the minimum $\Delta \chi^2$ is highest
around $\dmns^{\rm input}=90^\circ$
confirming the trend observed in \Fgref{oct.chi.wirct}.
We find from \Fgsref{oct.cont}(c) and \Fgvref{oct.cont}(d)
that the same trend holds even when the neutrino mass hierarchy
is inverted.
In all the four plots of \Fgref{oct.cont},
we recognize a high plateau around $\dmns^{\rm input}=90^\circ$
and a lower plateau around $\dmns^{\rm input}=-90^\circ$.
We can understand the trend by using the approximate expression
of the $\nu_\mu \to \nu_e$ transition probability, \eqref{ABe2}.
We first note that the $\nu_\mu \to \nu_e$ oscillation probability
is proportional to $(1+q)(1+A^e)\srct{2}$
around the oscillation maxima, $|\Delta_{13}=(2n+1)\pi|$,
When $q=-0.2$, the $\nu_e$ appearance rate is proportional to
$0.8(1+A^e)$.
In order to reproduce the same rate for $q=0.2$,
we should find a parameter set that makes the factor $1+A^e$
40\% smaller than its input value,
up to the uncertainty in $\srct{2}$,
which is assumed to be 0.01/$\srct{2}^{\rm input}$ in
\eqref{chi-para}.
This cannot be achieved for $\dmns^{\rm input}=90^\circ$,
because the input value of $1+A^e$ takes its minimum value.
On the other hand, if $\dmns^{\rm input}=-90^\circ$ the input
value of $1+A^e$ is large and it can be reduced siginificantly
by choosing $\dmns^{\rm fit}=90^\circ$ in the fit.
This explains why the minimum $\Delta \chi^2$ is larger around 
$\dmns^{\rm input}=90^\circ$ than that around
$\dmns^{\rm input}=-90^\circ$
when $\satm{}=0.4$ in \Fgsref{oct.cont}(a) and \Fgvref{oct.cont}(c).
When $q=0.2$, the same argument tells that we cannot compensate
for the large input value of $(1+q)(1+A^e)$
for $\dmns^{\rm input}=-90^\circ$.
This indeed explains the lower plateau around 
$\dmns^{\rm input}=-90^\circ$ observed in \Fgvref{oct.cont}(b)
and \Fgvref{oct.cont}(d).
The cause of the higher plateau around $\dmns^{\rm input}=90^\circ$
in these figures for $\satm{}=0.6$ is more subtle.
When $\dmns^{\rm input}=90^\circ$, 
$(1+A^e)^{\rm input}$ takes its smallest value,
and the reduction in $(1+q)$ from $(1+q)^{\rm input}=1+0.2$ to
$(1+q)^{\rm fit}=1-0.2$ can be compensated for by making 
$(1+A^e)^{\rm fit}$ larger by choosing $\dmns^{\rm fit} \simeq -90^\circ$.
This, however, necessarily makes the coefficient of
$|\Delta_{13}|/\pi$ in \eqref{Ae2} have the wrong sign,
and hence the ratio of the first peak $(|\Delta_{13}|\sim\pi)$
and the second peak $(|\Delta_{13}|\sim 3\pi)$
cannot be reproduced.
The higher plateau around $\dmns^{\rm input}=90^\circ$ in 
\Fgsref{oct.cont}(b) and \Fgvref{oct.cont}(d) for
$\satm{}^{\rm input}=0.6$,
and the lower plateau around $\dmns^{\rm input}=-90^\circ$ in 
\Fgsref{oct.cont}(a) and \Fgvref{oct.cont}(c) for
$\satm{}^{\rm input}=0.4$
can be explained as above.

\begin{figure}[p]
 \centering
  \includegraphics[scale=0.92]{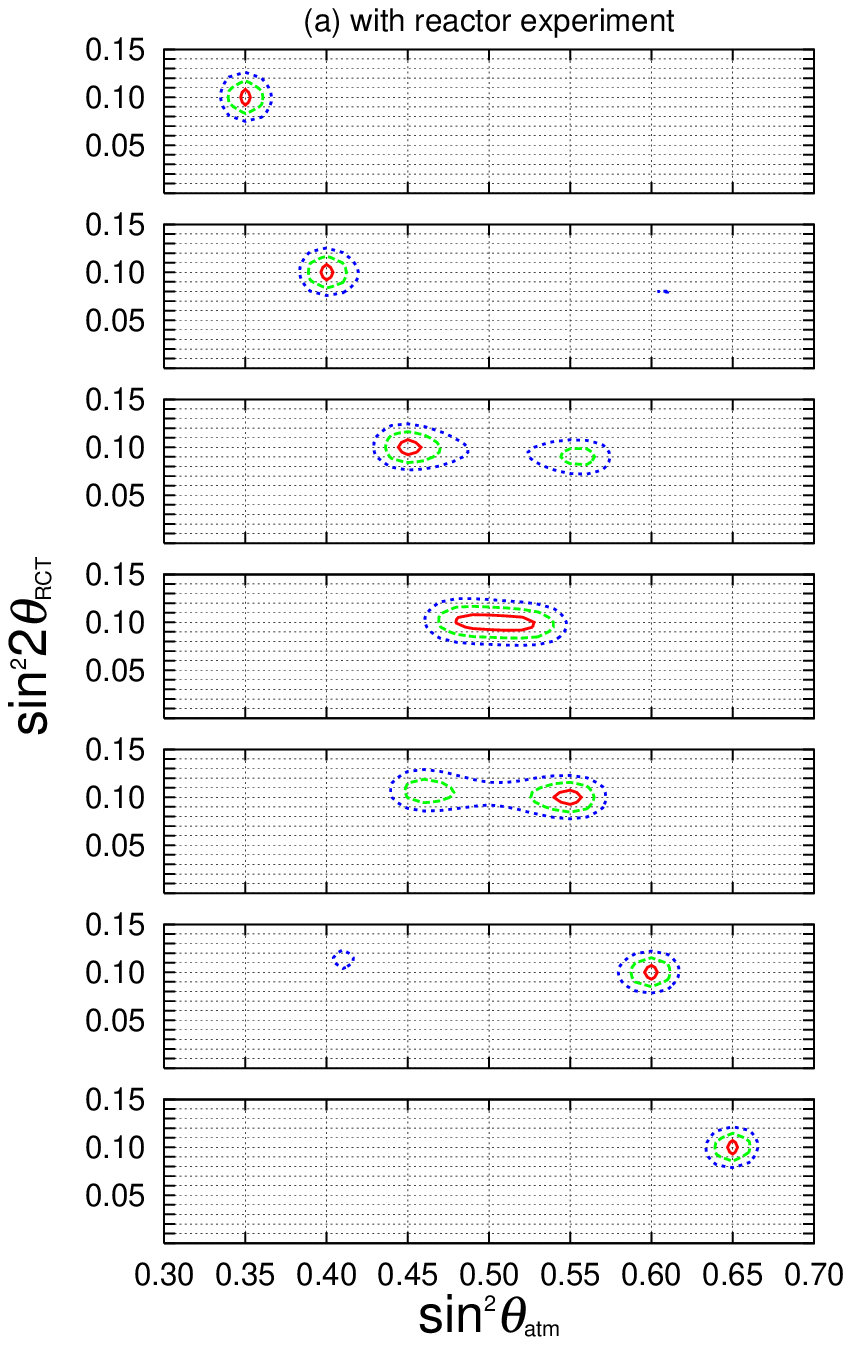}
~~~
  \includegraphics[scale=0.92]{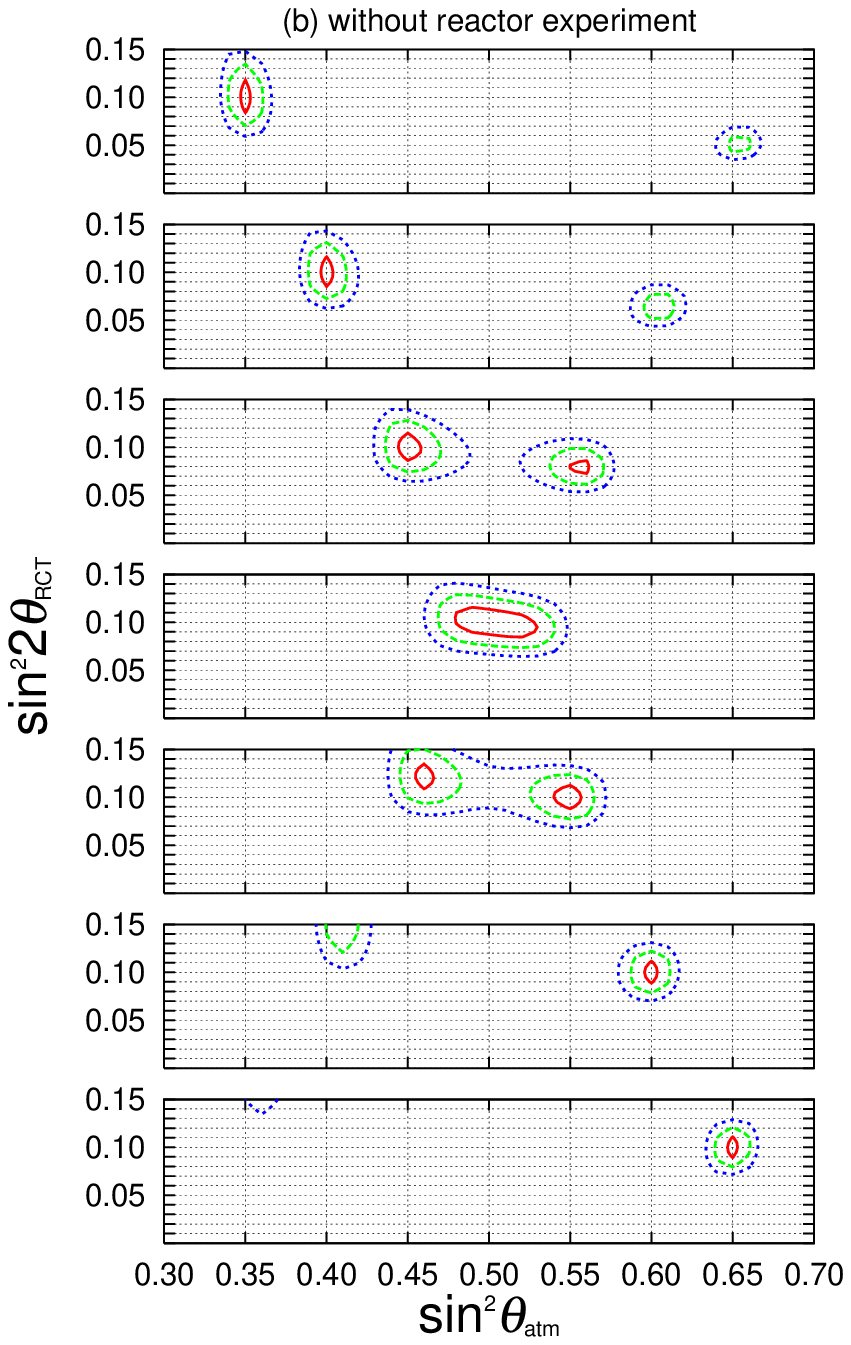}
 \caption{{%
The capability of the T2KK experiment for constraining the 
$\satm{}$ and $\srct{2}$.
Allowed regions in the plane of $\satm{}$ and $\srct{2}$
are shown for the same T2KK set up \Fgref{oct.chi.wirct}.
In each figure,
the event numbers are generated
at
$\dmns^{\rm input}=0^\circ$ and
$\srct{2}^{\rm input}=0.10$,
for
$\satm{}^{\rm input}=0.35$, 0.40, 0.45, 0.50, 0.55, 0.60, 0.65
from the 1st to the 7th row.
The other input parameters are the same as in \Fgref{oct.chi.wirct}.
In the left-hand-side plots, (a), we keep the constraint
on $\srct{2}$ form the future reactor experiment,
whereas in the right-hand-side plots, (b), 
we remove the external constraint on $\srct{2}$ in \eqref{chi-para}.
The $\Delta \chi^2=$ 1, 4, 9
contours are shown by the solid, dashed, and
dotted lines, respectively.
}}
\Fglab{oct.atm.rct}
\end{figure}

We show in \Fgref{oct.atm.rct}
the allowed region of $\satm{}$ and $\srct{2}$ by the T2KK experiment.
The event numbers are generated at
$\dmns^{\rm input}=0^\circ$ and
$\srct{2}^{\rm input}=0.10$
for
$\satm{}^{\rm input}=0.35$, 0.40, 0.45, 0.50, 0.55, 0.60, 0.65
from the 1st to the 7th row.
The other input parameters are the same as in \eqref{input_set}.
The allowed regions in the plane of $\satm{}$ and $\srct{2}$
are shown by the $\Delta \chi^2=$ 1, 4, 9 contours
depicted as solid, dashed, and dotted lines, respectively.
In the left-hand-side plots, (a), 
the constraint on $\srct{2}$ from the future reactor experiment
is kept in the $\Delta \chi^2$ function.
On the other hand, in the right-hand-side plots, (b),
the external constrant on $\srct{2}$ is removed from the
$\Delta \chi^2$ function in \eqref{chi-para}.
Comparing \Fgsref{oct.atm.rct}(a) and \Fgvref{oct.atm.rct}(b),
we find that the mirror solution around 
\begin{equation}
 \srct{2}^{\rm fit} = \dfrac{1+q}{1-q}\srct{2}^{\rm input}
\eqlab{mirror}
\end{equation}
cannot
be excluded without the information from the future reactor experiment.

\begin{figure}[t]
\centering
  \includegraphics[scale=0.6]{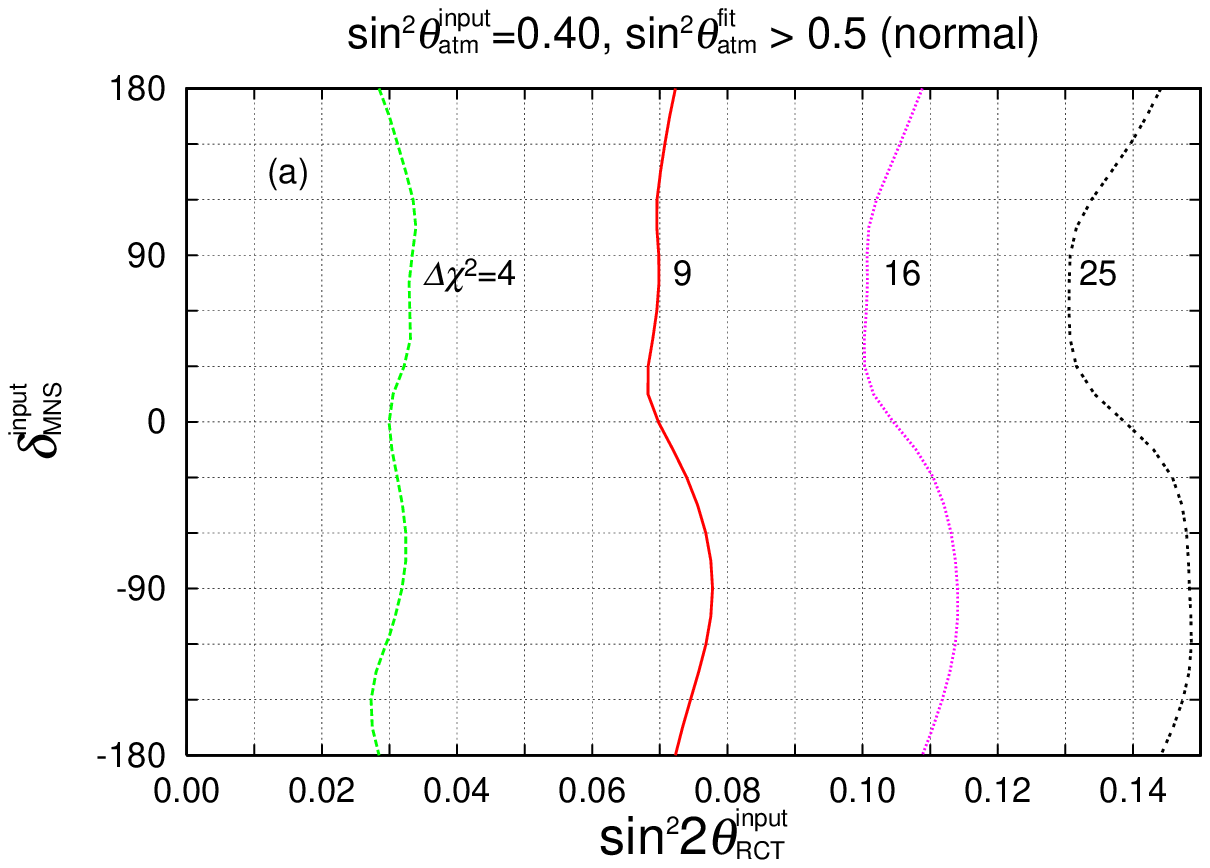}
~~~
  \includegraphics[scale=0.6]{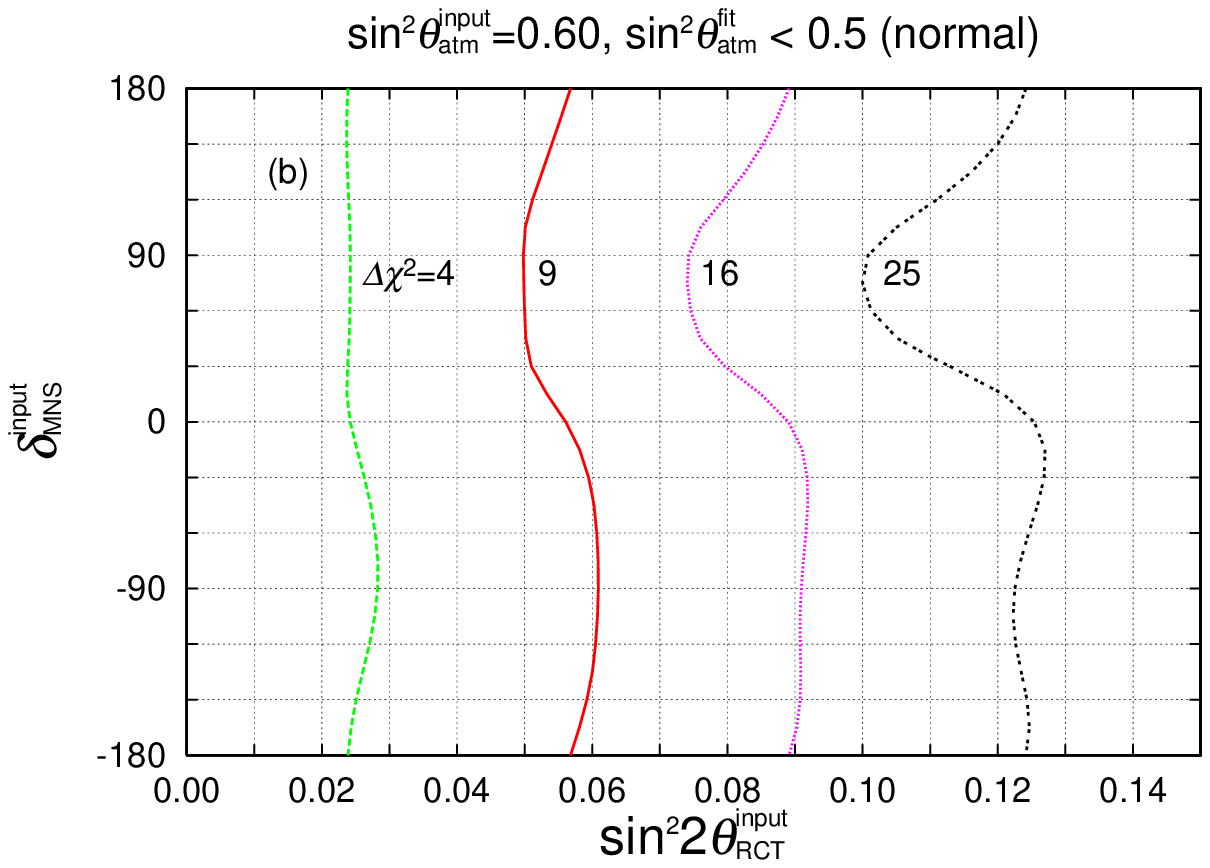}

  \includegraphics[scale=0.6]{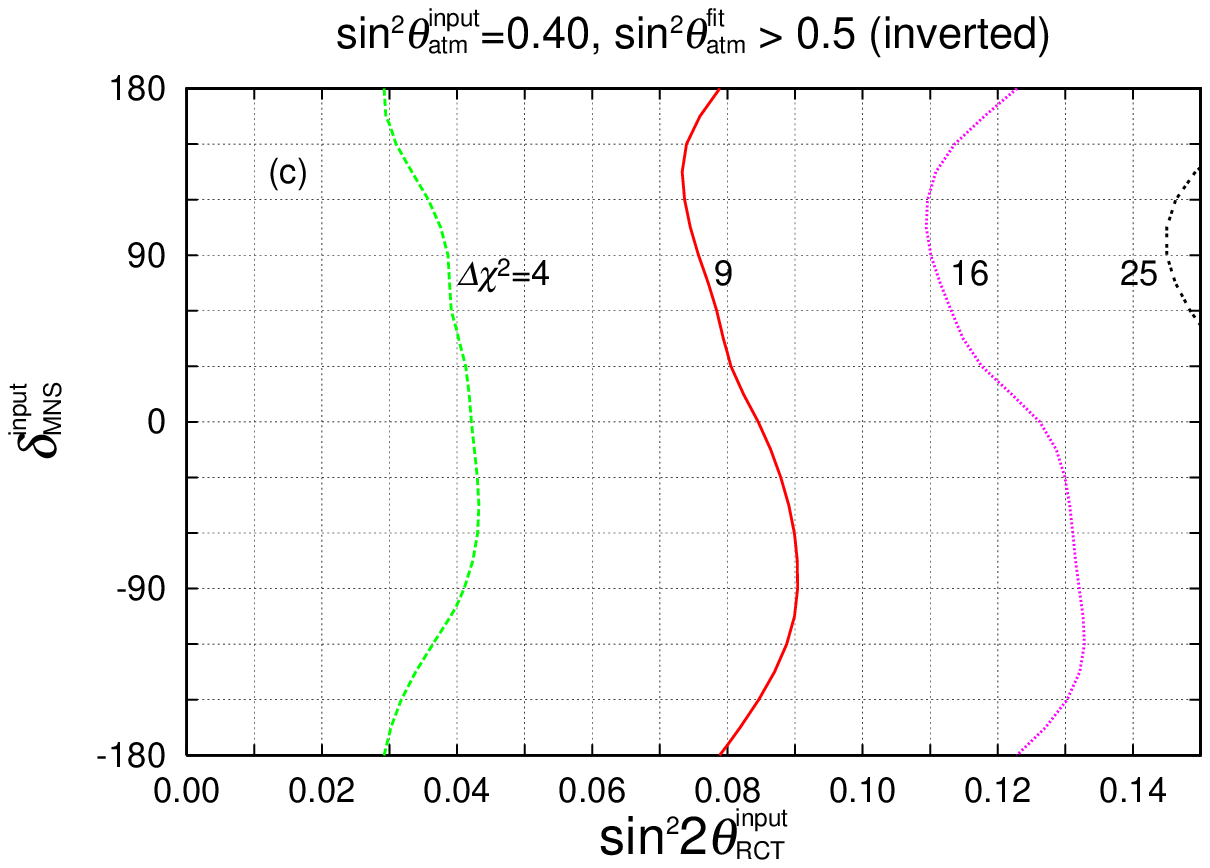}
~~~
  \includegraphics[scale=0.6]{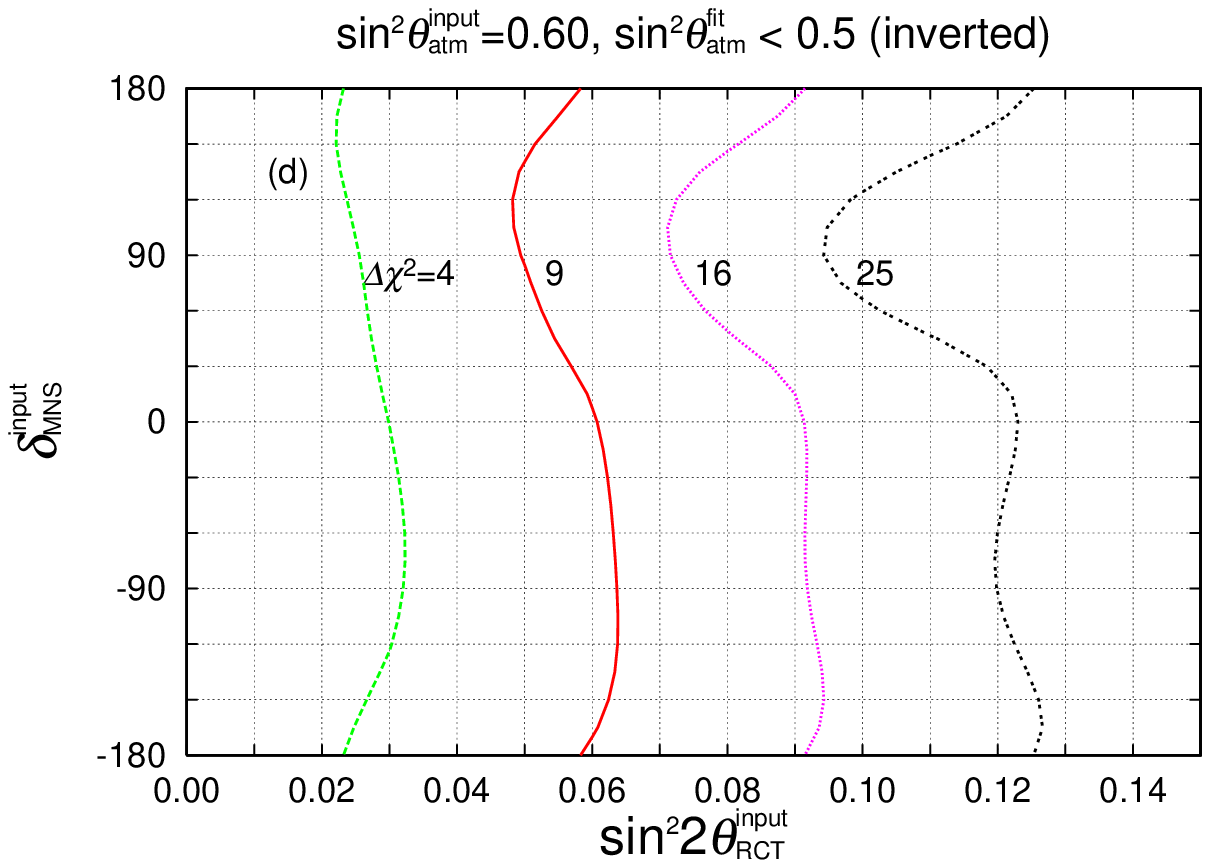}
 \caption{{%
The same as \Fgref{oct.cont},
but with 5 times larger exposure ($\numt{25}{21}$ POT).
}}
\Fglab{oct.contH}
\end{figure}
Before closing the section, we examine the impact of upgrading
the J-PARC beam intensity be a factor of 5 \cite{upgrade}
on the resolution of the octant degeneracy.
Such an upgrade is desirable especially if the neutrino mass hierarchy
is inverted, because the octant degeneracy between 
$\satm{}=0.4$ and 0.6 cannot be resolved at $3\sigma$ unless
$\dmns \simeq 90^\circ$; see \Fgsref{oct.cont}(c) and \Fgvref{oct.cont}(d).

We show in \Fgref{oct.contH} the same contour plots as in
\Fgref{oct.cont},
but with 5 times larger exposure
($\numt{25}{21}$ POT).
It is found that the degeneracy between $\satm{}=0.4$ and 0.6 can now 
be resolved at $3\sigma$ level for
$\srct{2}>0.08$(0.09), when the hierarchy is normal (inverted).
Comparing \Fgref{oct.cont} and \Fgref{oct.contH},
however, 
we find that the sensitivity does not improved as much as we would
hope with 5 times higher statistics.
The minimum $\Delta \chi^2$ value does not grow by a factor 5,
because the capability of resolving the octant degeneracy is now
dictated by the accuracy of the external constraint on $\srct{2}$
from the future reactor experiment,
\begin{equation}
 \dfrac{\delta\srct{2}}{\srct{2}} = \dfrac{0.01}{\srct{2}}\,,
\eqlab{delta-srct}
\end{equation}
which we assume in \eqref{chi-para}.
The fractional uncertainty of $\srct{2}$ is 10\% for $\srct{2}=0.1$,
but it is 17\% for $\srct{2}=0.06$.
If $\srct{2}$ turns out to be even smaller, the fractional error
grows and the mirror solution \eqref{mirror} can no more be resolved.
If $\srct{2}$ turns out to be smaller than 0.06, further reduction
of its error in the future experiments with reactor and/or
the beta beam \cite{beta}.

\section{Mass hierarchy and the octant degeneracy}
\label{sec:mass}
In this section, we examine the effect of the octant
degeneracy on the capability of the T2KK experiment
to determine the neutrino mass hierarchy pattern.

\begin{figure}[t]
\centering
  \includegraphics[scale=0.6]{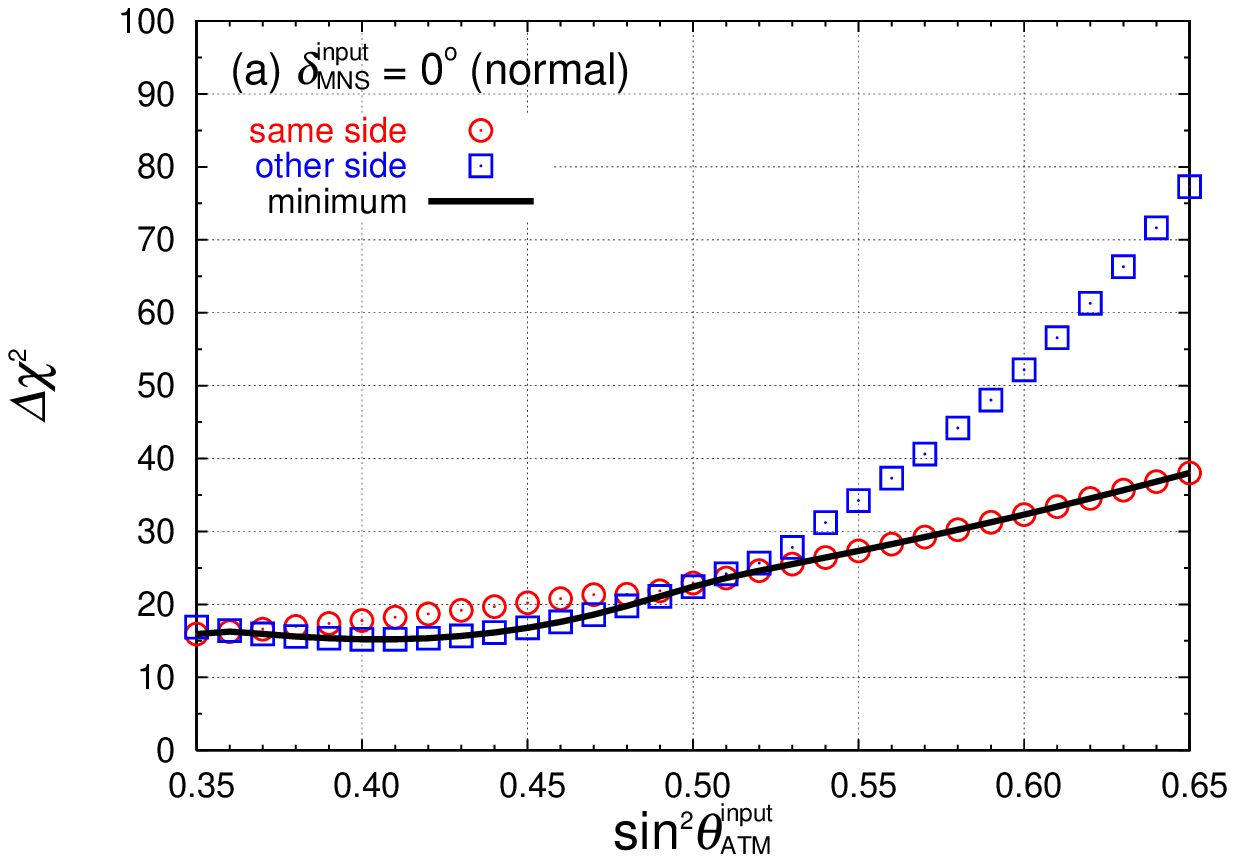}
~~~~~
  \includegraphics[scale=0.6]{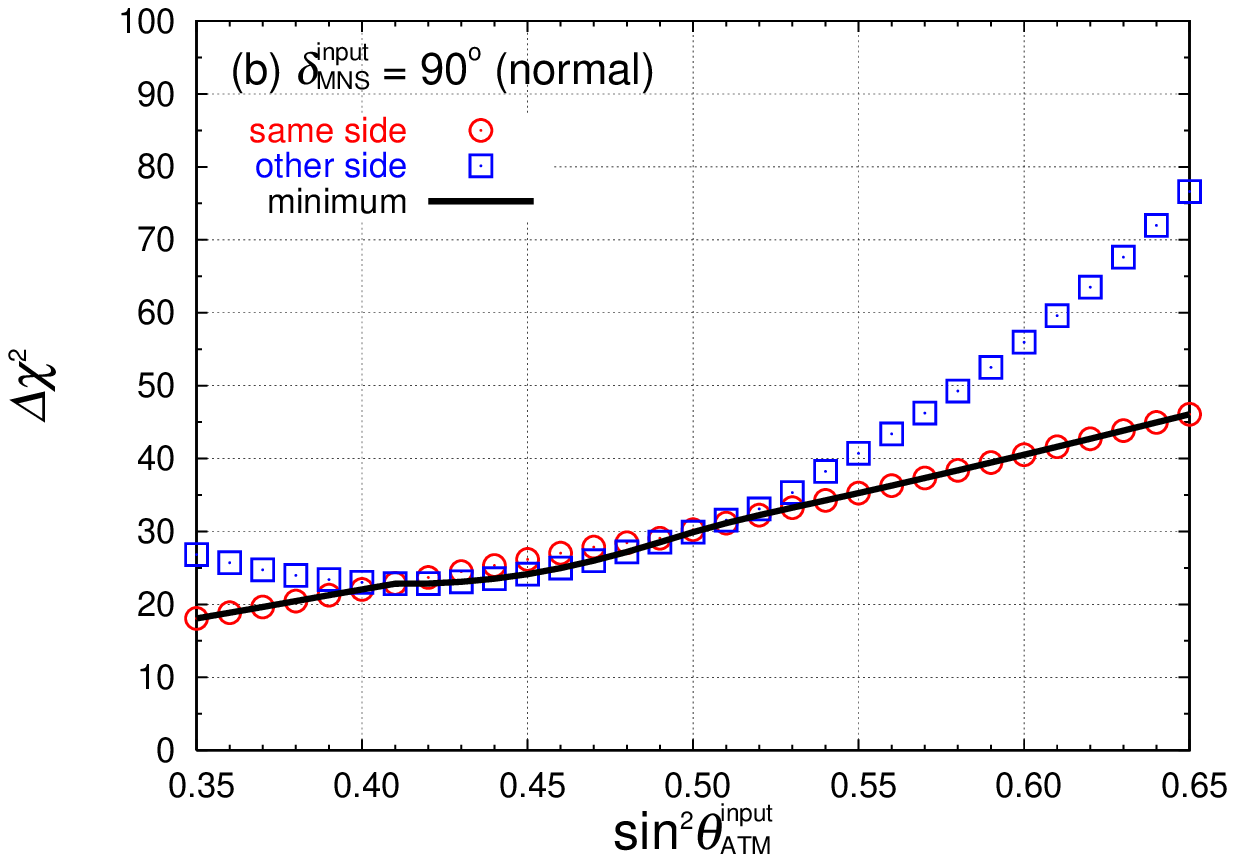}

  \includegraphics[scale=0.6]{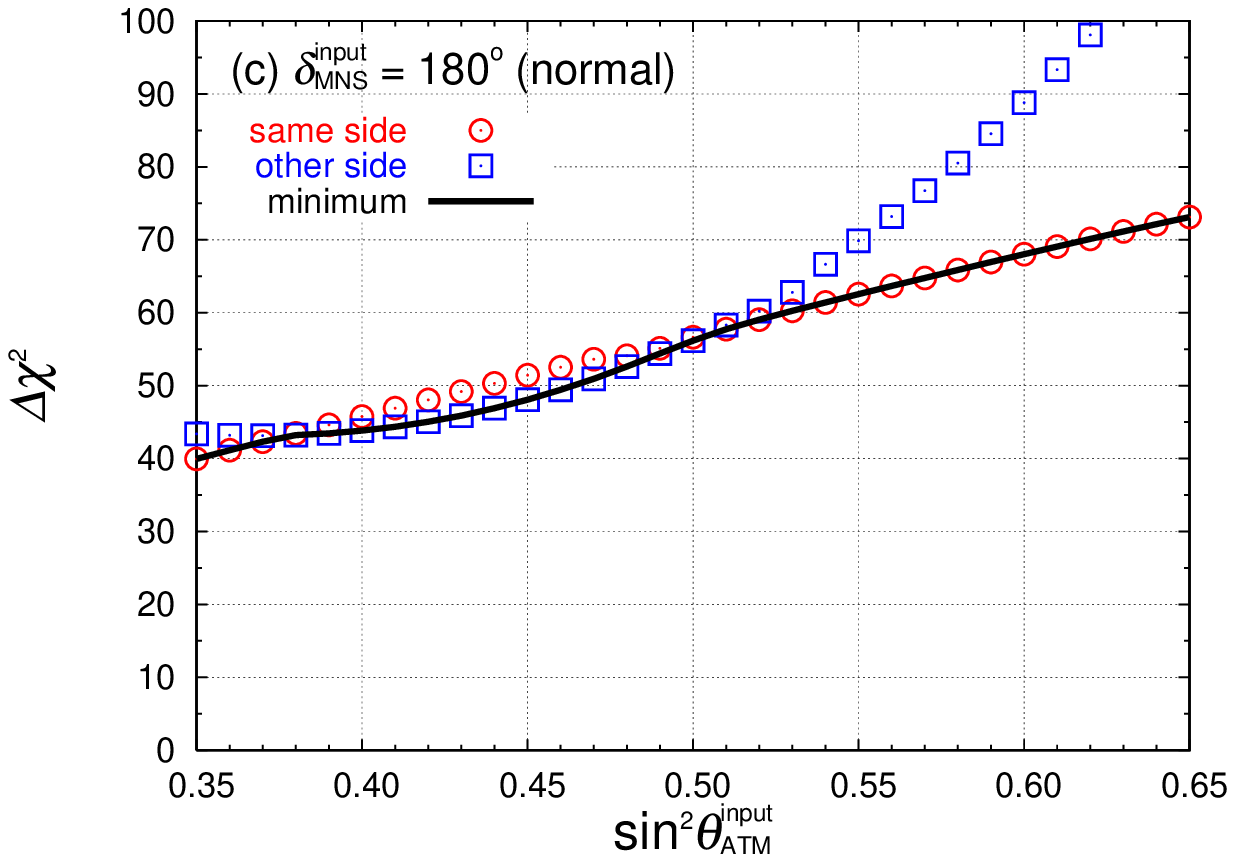}
~~~~~
  \includegraphics[scale=0.6]{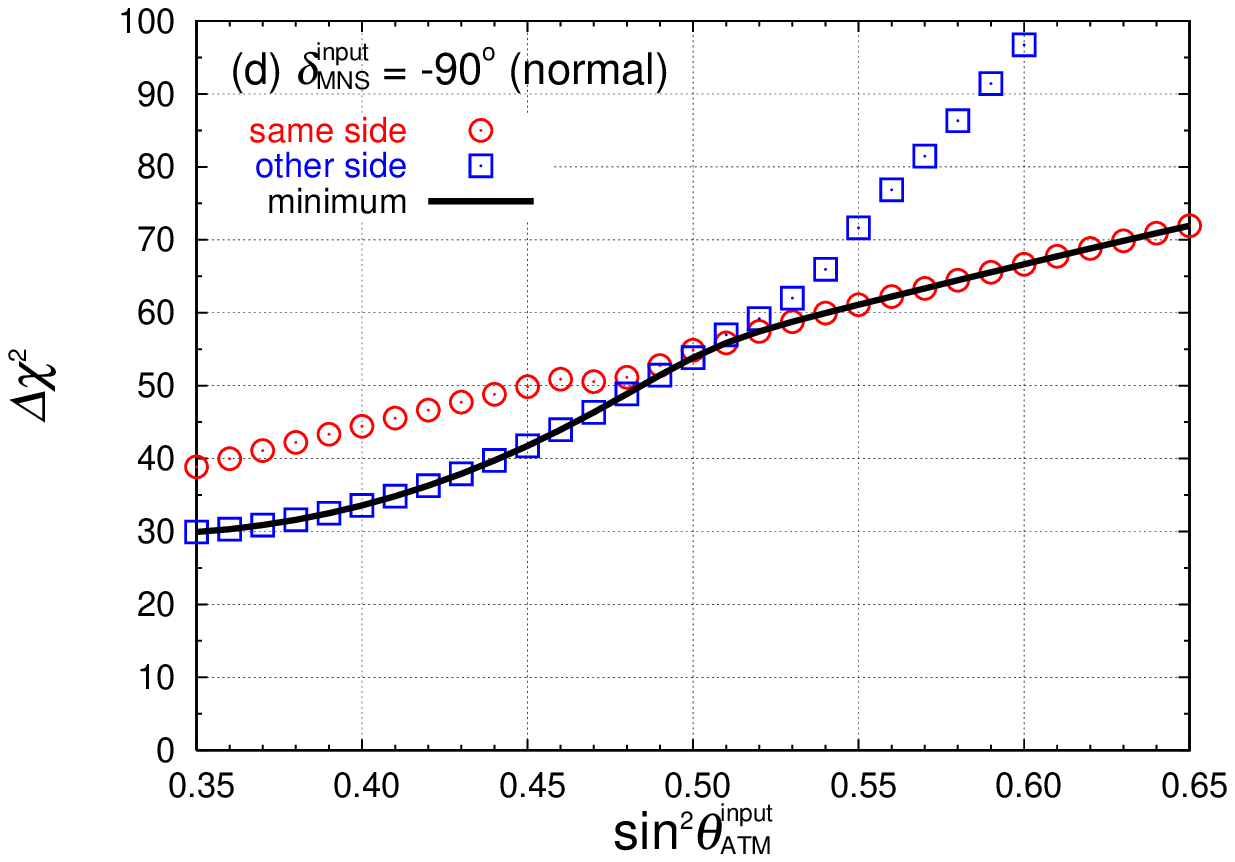}
 \caption{{%
 Minimum $\Delta \chi^2$ as a function of $\satm{}^{\rm input}$
for the same T2KK setting as in \Fgref{oct.chi.wirct}.
In each figure, the event numbers are calculated 
for the parameters of \eqref{input_set} with
$\dmns^{\rm input}=0^\circ$ (a),
$90^\circ$ (b),
$180^\circ$ (c),
$-90^\circ$ (d),
under the normal hierarchy,
and the fit has been performed by assuming the inverted hierarchy.
The solid line gives the minimum $\Delta \chi^2$.
The open circle (square) denotes the minimum value of
$\Delta \chi^2$ when
the sign of $q^{\rm input} q^{\rm fit}$ is positive (negative).
}}
\Fglab{chi.mass.nor}
\end{figure}
\Figref{chi.mass.nor} shows 
the minimum $\Delta \chi^2$ as a function of
$\satm{}^{\rm input}$ for the T2KK experiment to
determine the mass hierarchy pattern
with the same OAB combination of \Fgref{oct.chi.wirct}.
In each figure, the event numbers are calculated for
$\dmns^{\rm input}=0^\circ$ (a),
$90^\circ$ (b),
$180^\circ$ (c), and
$-90^\circ$ (d), 
when the normal hierarchy is realized.
The other parameters are listed in \eqref{input_set}.
The fit has been performed by surveying the whole parameter
space by assuming the wrong hierarchy.
The solid line gives the minimum $\Delta \chi^2$.
The open circle denotes the minimum $\Delta \chi^2$
when the sign of $q^{\rm input} q^{\rm fit}$, or that of 
$(1-2\satm{}^{\rm input})(1-2\satm{}^{\rm fit})$,
is positive,
whereas the open square gives the minimum $\Delta \chi^2$
when $q^{\rm input} q^{\rm fit}$ is negative.

When $q^{\rm fit}$ takes the same sign as $q^{\rm input}$,
$\satm{}^{\rm fit} \sim \satm{}^{\rm input}$ is favored,
and
the reduction of the $\nu_\mu \to \nu_e$ oscillation amplitude
$(1+A^e)$ in \eqref{Ae2} for the inverted hierarchy, 
$\Delta_{13} \sim -\pi$, cannot be compensated for
in the two detector experiment \cite{HOS1, HOS2}.
Because the $\nu_\mu \to \nu_e$ rate is proportional to
$\satm{}^{\rm input}$,
the resulting increase in the discrepancy leads 
to the linear dependence of the minimum $\Delta \chi^2$
on $\satm{}^{\rm input}$ observed for the open 
circle points.
On the other hand, when $\satm{}^{\rm input}<0.5$ $(q^{\rm input}<0)$,
it is possible to compensate for the reduction of the $1+A^e$ factor
of the $\nu_\mu \to \nu_e$ oscillation amplitude by choosing
$q^{\rm fit} \sim -q^{\rm input}$, since 
$\satm{}^{\rm fit}= \satm{}^{\rm input}
(1+q^{\rm fit})/(1+q^{\rm input}) > 0.5$.
This explains why the open square points for 
$q^{\rm input} q^{\rm fit} <0$ gives the lowest $\Delta \chi^2$
for $\satm{}^{\rm input}<0.5$.
When $\satm{}^{\rm input}$ is significantly lower than 0.5,
however, the enlargement factor of 
$\satm{}^{\rm fit}/\satm{}^{\rm input} = (1+q^{\rm fit})/(1+q^{\rm input})$
overshoots the reduction due to the matter effect,
especially for the $\nu_\mu \to \nu_e$ rate at SK when
the matter effect is small.
When $\dmns^{\rm input}=-90^\circ$, shown in \Fgref{chi.mass.nor}(d),
this reduction in the minimum $\Delta \chi^2$ by using the octant
degeneracy is most significant
because the overshooting of the $\nu_\mu \to \nu_e$ rate can be
partially compensated by choosing $\sin \dmns^{\rm fit}>0$.

In \Fgref{chi.mass.inv}, we show the minimum $\Delta \chi^2$ as
a function of $\satm{}^{\rm input}$ when the neutrino mass hierarchy
is inverted.
The gradual increase of the $\Delta \chi^2$ as $\satm{}^{\rm input}$
grows can also be seen for the open circle points where the fit is
restricted to the parameter space that satisfies
$q^{\rm fit} q^{\rm input}>0$.
This reflects the increase of the $\nu_\mu \to \nu_e$ event rate as
$\satm{}^{\rm input}$ increases,
independent of the hierarchy pattern.
On the other hand, the minimum $\Delta \chi^2$ 
for the parameter space of $q^{\rm fit} q^{\rm input}<0$,
plotted by open squares,
gives the lowest $\Delta \chi^2$ when
$\satm{}^{\rm input} > 0.5$.
This is because the reduction of the $\nu_\mu \to \nu_e$ rate
in the fit, which is proportional to 
$\satm{}^{\rm fit}/\satm{}^{\rm input}=(1+q^{\rm fit})/(1+q^{\rm input})<1$,
for $q^{\rm fit} < 0 <q^{\rm input}$,
can compensate for the reduction due to the matter effect
when the hierarchy is inverted.
The reduction of the minimum $\Delta \chi^2$ due to the octant
degeneracy is strong at $\satm{}^{\rm input}>0.5$ for the 
inverted hierarchy, and the increased sensitivity to the 
mass hierarchy pattern for large $\satm{}$ is lost 
when it is inverted.

\begin{figure}[t]
 \centering
  \includegraphics[scale=0.6]{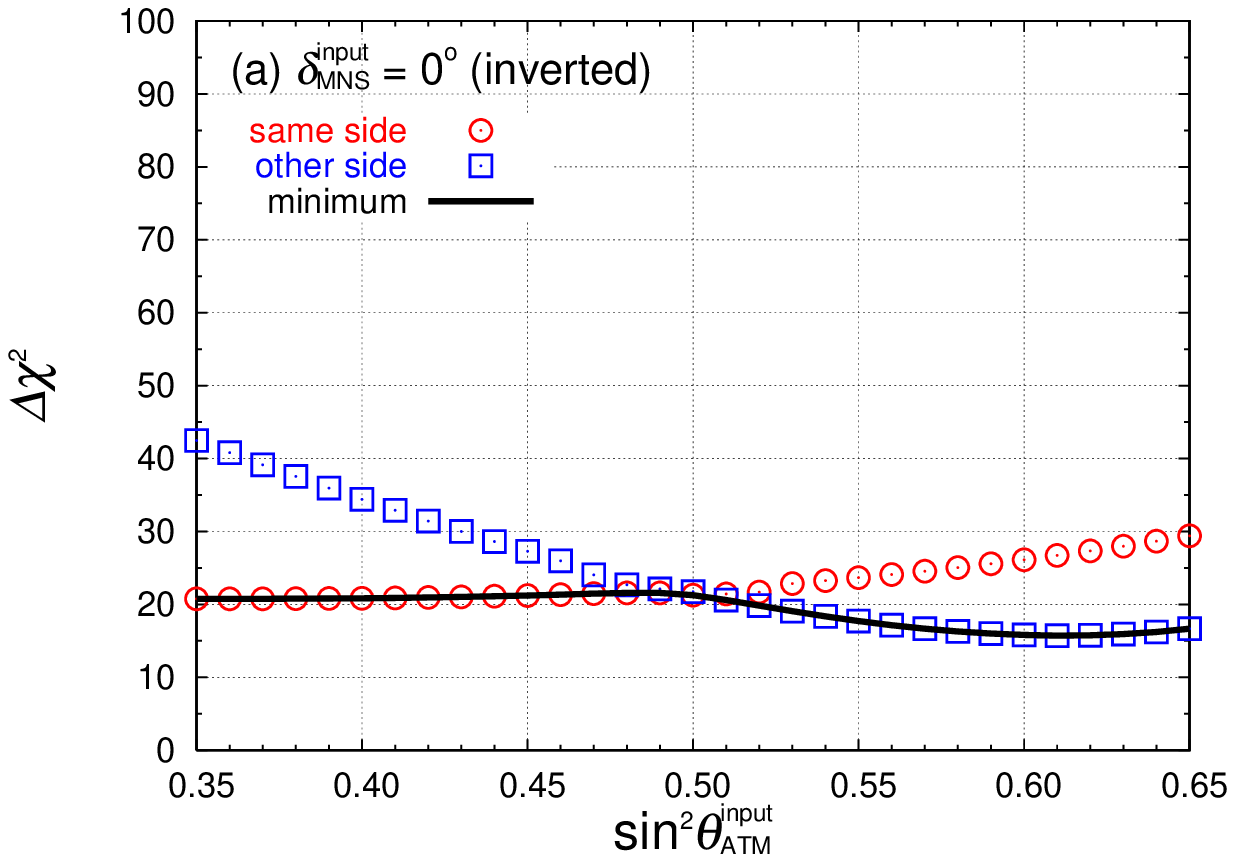}
~~~~~
  \includegraphics[scale=0.6]{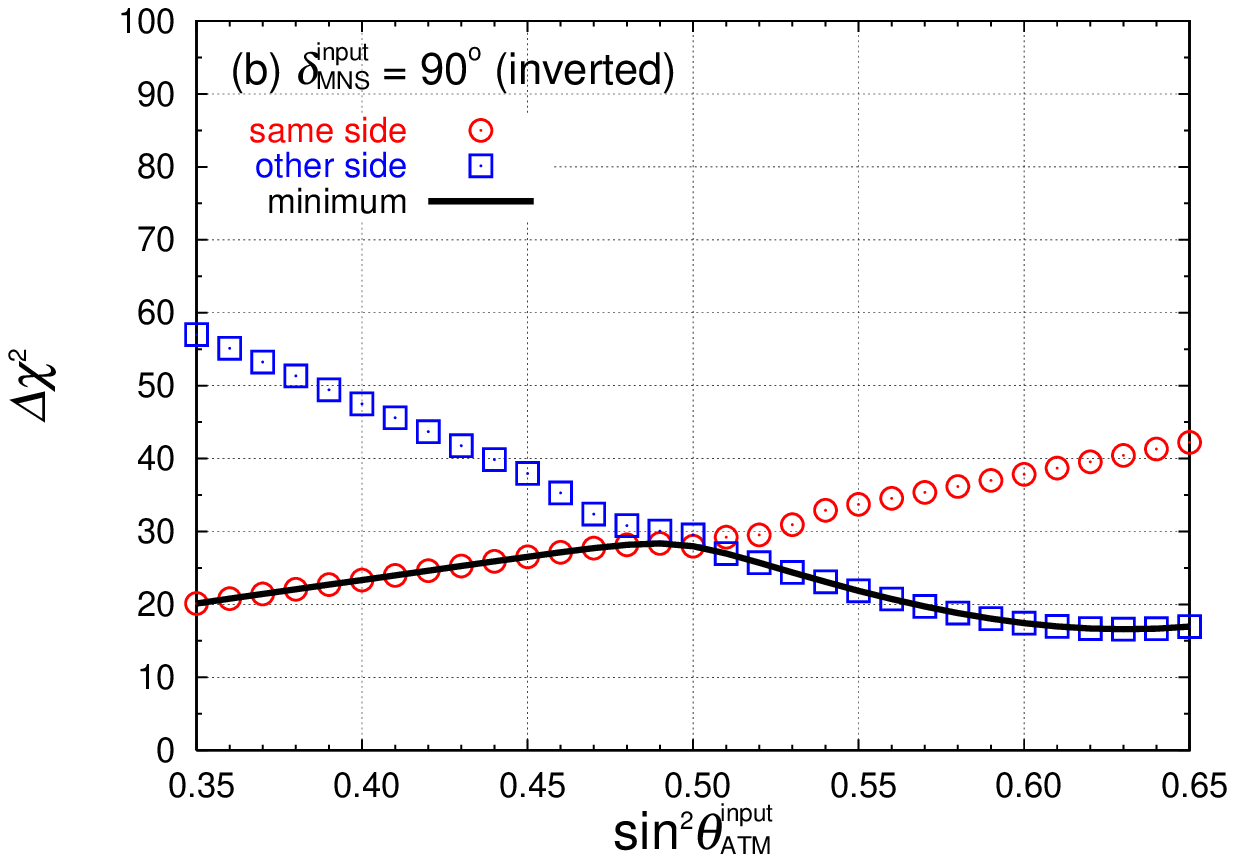}

  \includegraphics[scale=0.6]{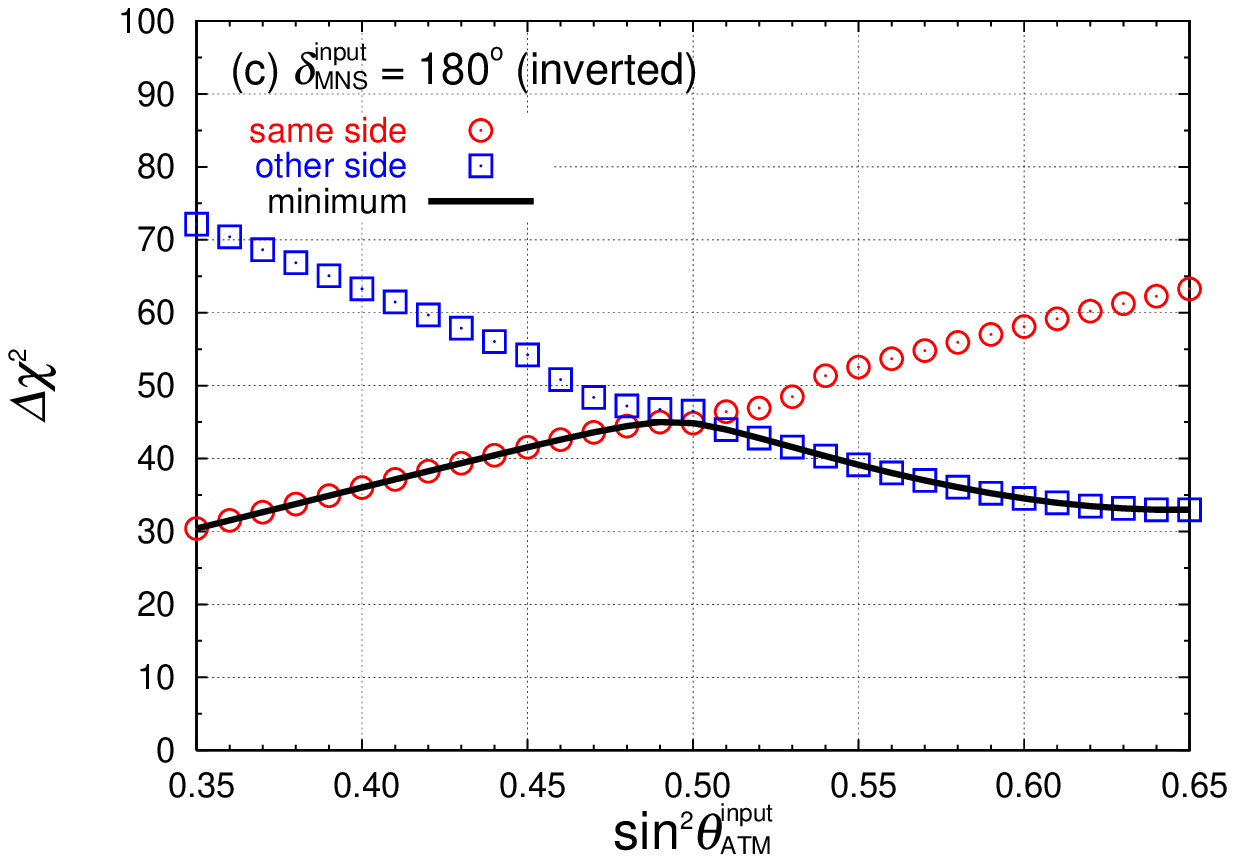}
~~~~~
  \includegraphics[scale=0.6]{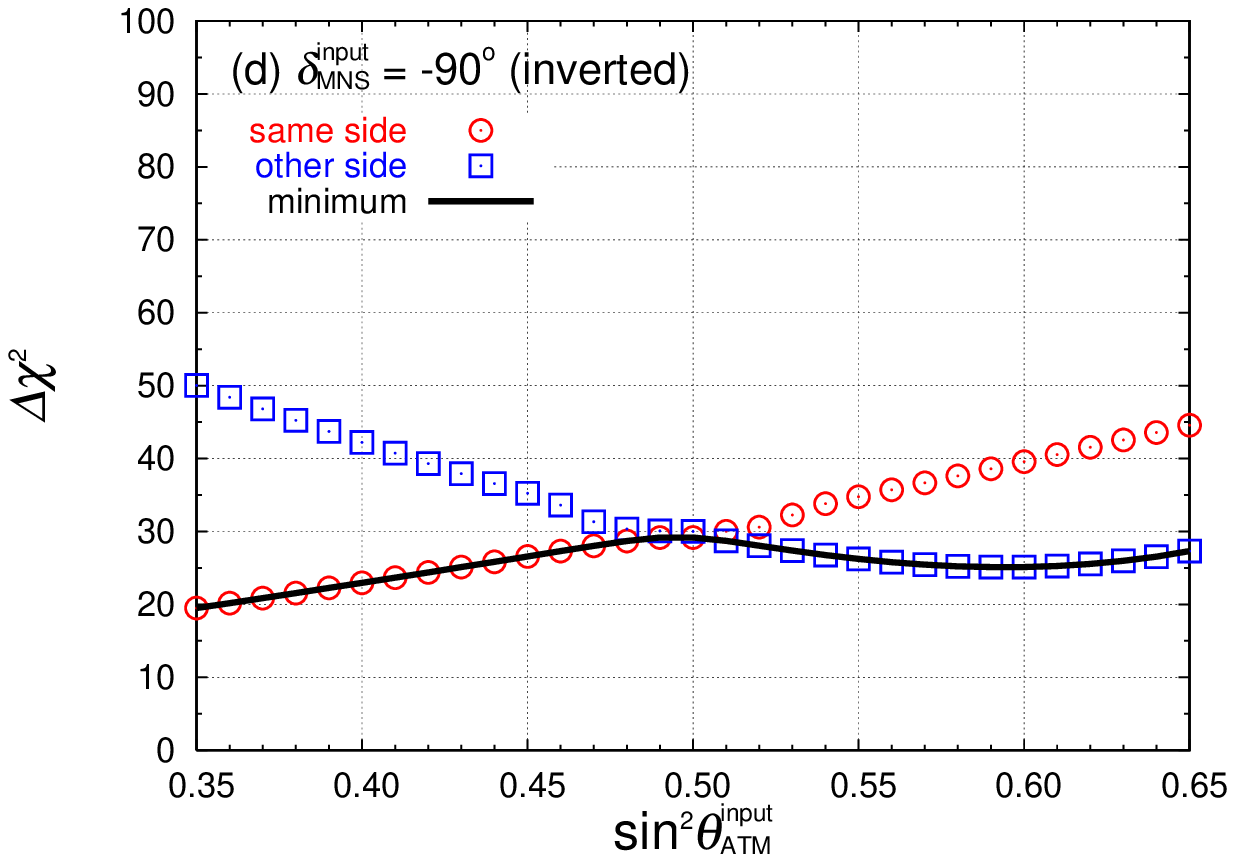}
 \caption{{%
The same as \Fgref{chi.mass.nor}, but
when the inverted hierarchy is realized in nature
and the fit is performed by assuming the normal hierarchy.
}}
\Fglab{chi.mass.inv}
\end{figure}

\begin{figure}[p]
\centering
  \includegraphics[scale=0.6]{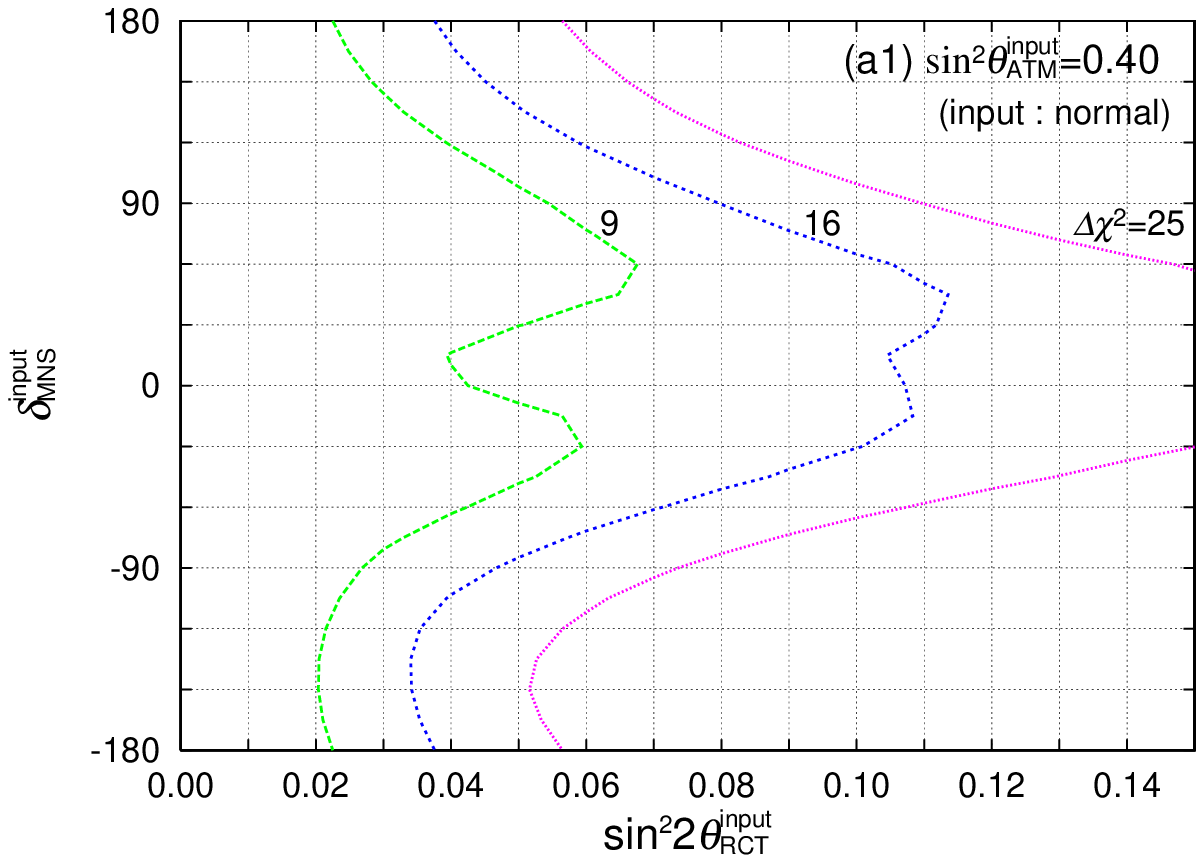}
  \includegraphics[scale=0.6]{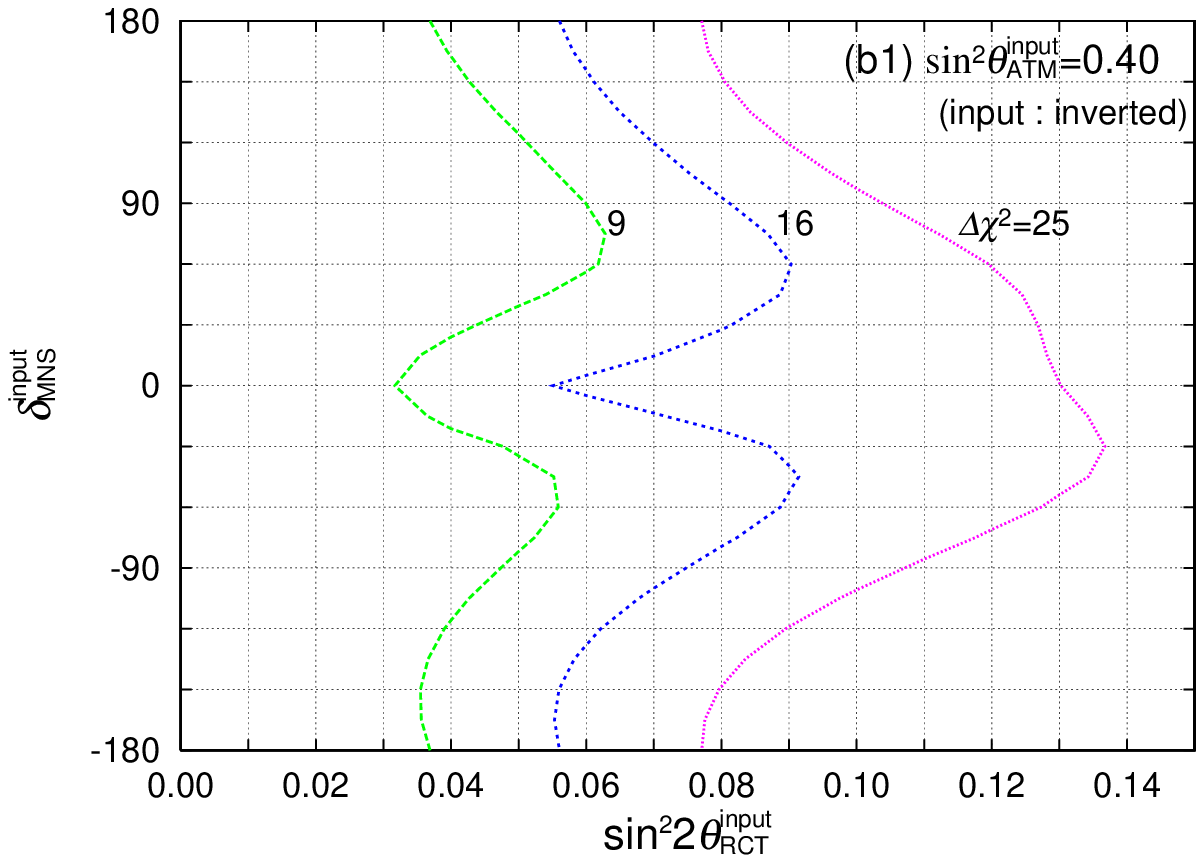}

  \includegraphics[scale=0.6]{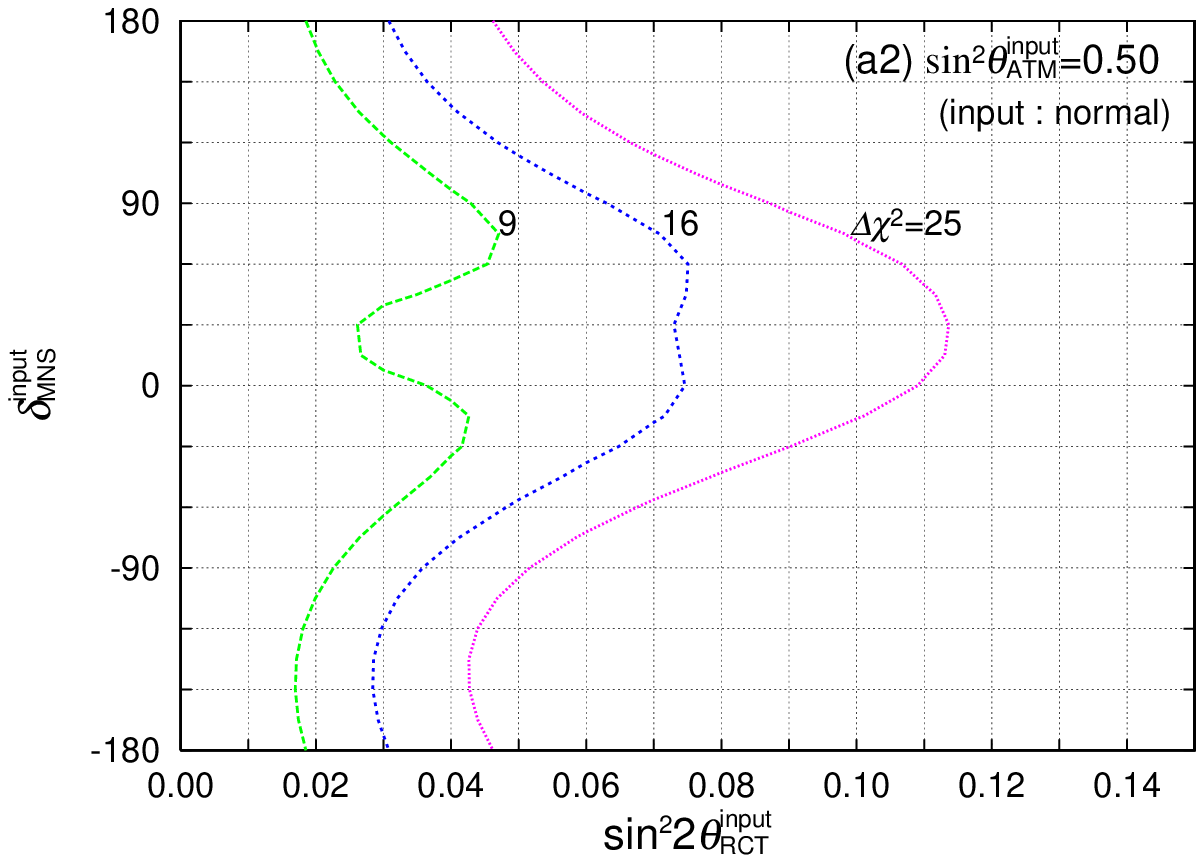}
  \includegraphics[scale=0.6]{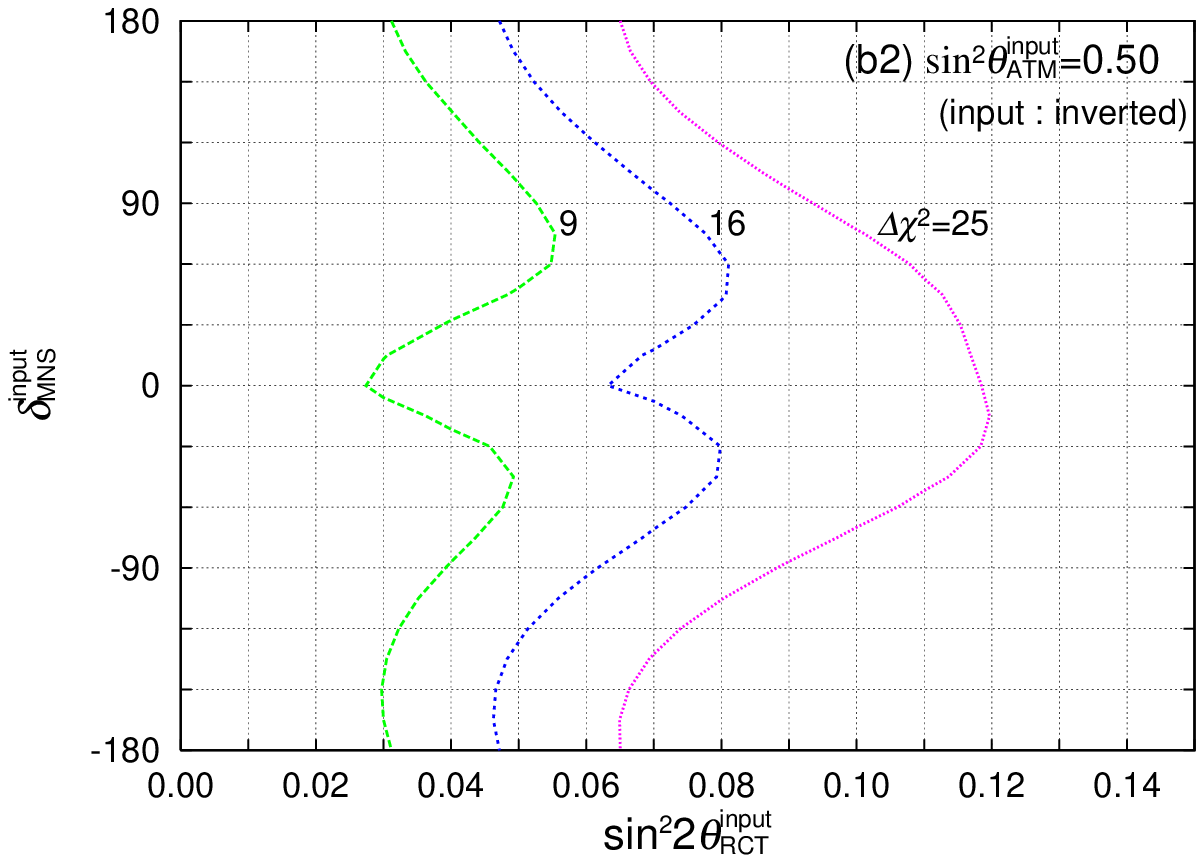}

  \includegraphics[scale=0.6]{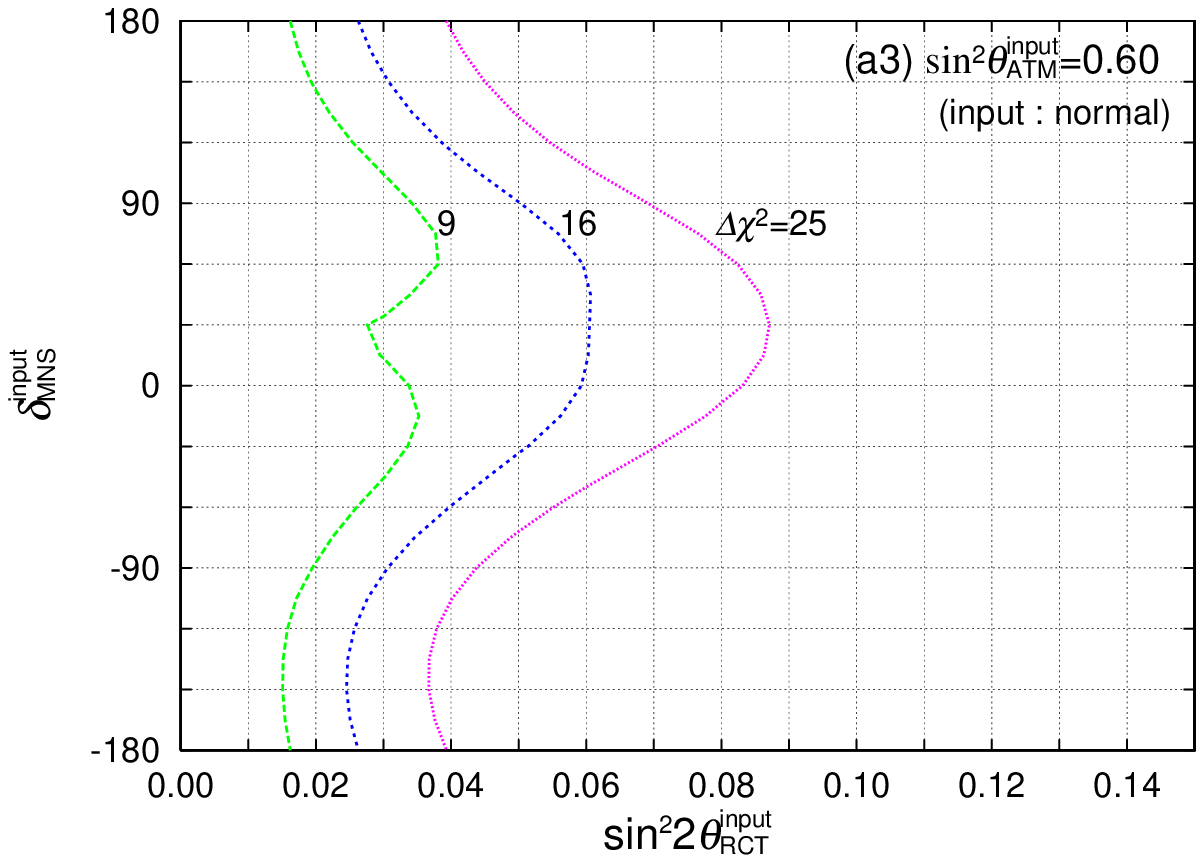}
  \includegraphics[scale=0.6]{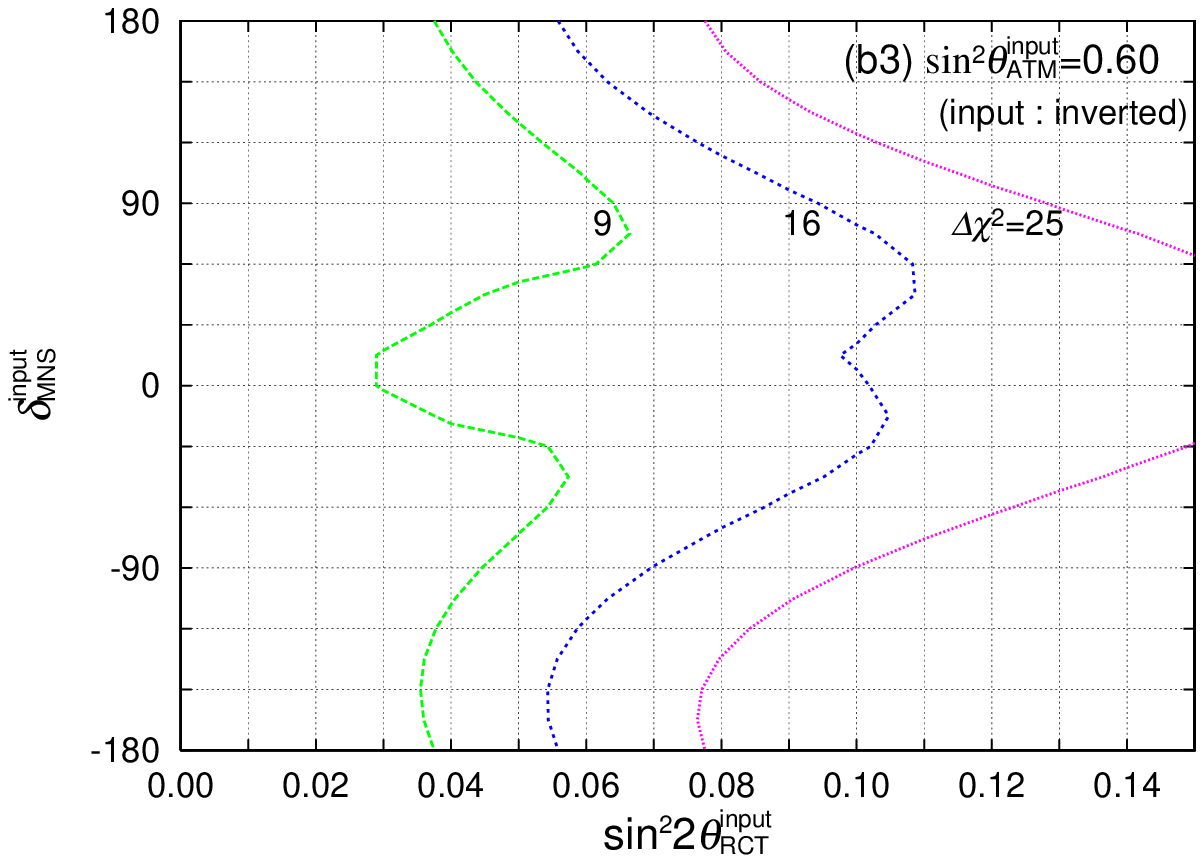}
 \caption{{%
The potential of the T2KK experiment to determine the neutrino mass
hierarchy.
 The input data calculated for the normal hierarchy (a1, a2, a3)
or for the inverted hierarchy (b1, b2, b3),
and the fit has been performed by assuming the wrong hierarchy.
In each figure, the minimum $\Delta \chi^2$ is obtained for the input
data calculated at various 
($\srct{2}^{\rm input}$, $\dmns^{\rm input}$)
with 
$\satm{}^{\rm input}=0.40$ for (a1, b1),
$\satm{}^{\rm input}=0.50$ for (a2, b2),
or
$\satm{}^{\rm input}=0.60$ for (a3, b3).
The other input parameters are the same as those of
\Fgref{oct.chi.wirct}.
The contours of the minimum $\Delta \chi^2$ are
shown for $\Delta \chi^2=9$, 16, 25.
}
\Fglab{chi.cont.mass}
}
\end{figure}
In \Fgref{chi.cont.mass},
we show the capability of the T2KK experiment 
to determine the mass hierarchy pattern as contour plots
of the minimum $\Delta \chi^2$ value on the parameter space
of $\srct{2}^{\rm input}$ and $\dmns^{\rm input}$.
In each figure, 
the input date are calculated for the model parameters
at various $\srct{2}^{\rm input}$ and $\dmns^{\rm input}$
values,
with
$\satm{}^{\rm input}=0.40$ in (a1, b1),
$\satm{}^{\rm input}=0.50$ in (a2, b2),
or
$\satm{}^{\rm input}=0.60$ in (a3, b3).
The left-hand figures (a1, a2, a3) are for the
the normal hierarchy,
and
the right-hand figures (b1, b2, b3) are for the
the inverted hierarchy.
The other input parameters are the same as those of \Fgref{chi.mass.nor},
and in \eqref{input_set}.
All the fit parameters are varied freely to minimize the 
$\Delta \chi^2$ function, under the constraint of
the opposite mass hierarchy.
The resulting values of minimum $\Delta \chi^2$ are
shown as contours for $\Delta \chi^2=9$, 16, 25.
The contours of \Fgsref{chi.cont.mass}(a2) and \Fgvref{chi.cont.mass}(b2)
are identical to those of Fig.6 of Ref.~\cite{HOS2},
which we copy for the purpose of comparison.

It is clearly seen in \Fgref{chi.cont.mass} that
the main feature of the T2KK ability for the mass hierarchy determination
at $\satm{2}^{\rm input}=0.96$, $\satm{}=0.4$ (a1, b1) or 0.6 (a3, b3),
are not much different from those at $\satm{2}^{\rm input}=1.0$
(a2, b2),
such as the fact the minimum $\Delta \chi^2$
around $\dmns^{\rm input}\simeq 0^\circ$
is smaller than that around $\dmns^{\rm input}\simeq 180^\circ$.
Close examination of \Fgref{chi.cont.mass}, however,
reveals the followings.
In case of the normal hierarchy, $m^2_3-m^2_1>0$,
the minimum $\Delta \chi^2$ grows with growing $\satm{}^{\rm input}$,
and the mass hierarchy can be determined at $3\sigma$ level 
for $\srct{2}^{\rm input} \gsim 0.07$ if $\satm{}^{\rm input}=0.40$ (a1),
whereas the same holds
for $\srct{2}^{\rm input} \gsim 0.04$ if $\satm{}^{\rm input}=0.60$
(a3).
This is mainly because the $\nu_\mu \to \nu_e$ event rate grows with
$\satm{}^{\rm input}$ and because the presence of the
octant degeneracy does not disturb the measurement significantly,
as can be seen from the open circle points in \Fgref{chi.mass.nor}.
In contract, no significant improvement in the hierarchy
discrimination power
is found for $\satm{}^{\rm input} > 0.5$ in case of
the inverted hierarchy.
This is because the octant degeneracy between $\satm{}^{\rm fit}=0.6$
and $\satm{}^{\rm fit}=0.4$ allows us to compensate for the matter effect
reduction of the $\nu_\mu \to \nu_e$ rate.
We find that the best hierarchy discrimination is achieved at 
$\satm{}^{\rm input}=0.5$ for all $\dmns$ values, confirming the
trends of \Fgref{chi.mass.inv}.

Summing up, the T2KK two detector experiments can resolve the mass
hierarchy pattern in the presence of the octant degeneracy.
If the hierarchy is normal, the discriminating power grows with
increasing $\satm{}^{\rm input}$.
On the other hand, 
if the hierarchy is inverted, the discriminating power reduces
both at $\satm{}^{\rm input}<0.5$ and at $\satm{}^{\rm input}>0.5$:
it reduces at $\satm{}<0.5$ because of the lower rate of the 
$\nu_\mu \to \nu_e$ events,
while it reduces at $\satm{}>0.5$ because of the octant degeneracy.
\section{CP phase and the octant degeneracy}
\label{sec:CP}
In this section, 
we investigate the relation between the CP phase measurement
and the octant degeneracy.

\begin{figure}[t]
 \centering
  \includegraphics[scale=0.6]{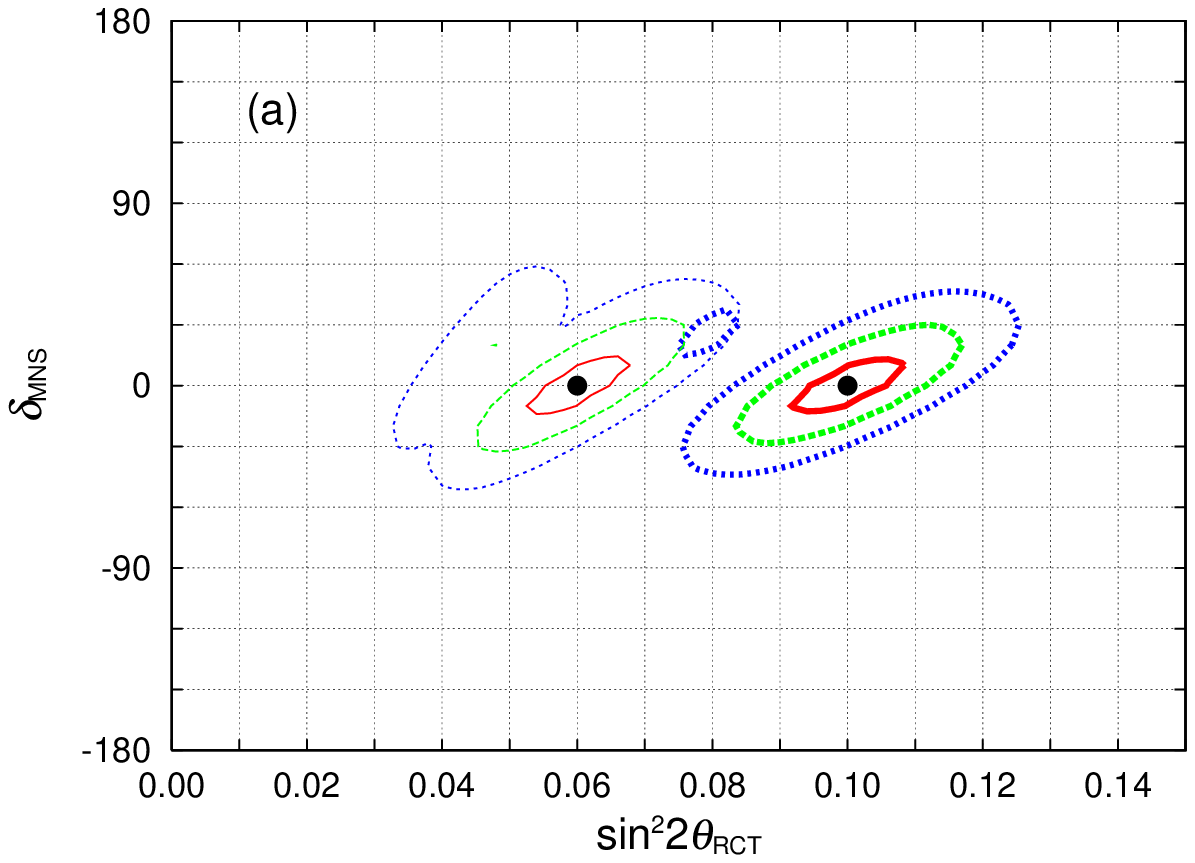}
  \includegraphics[scale=0.6]{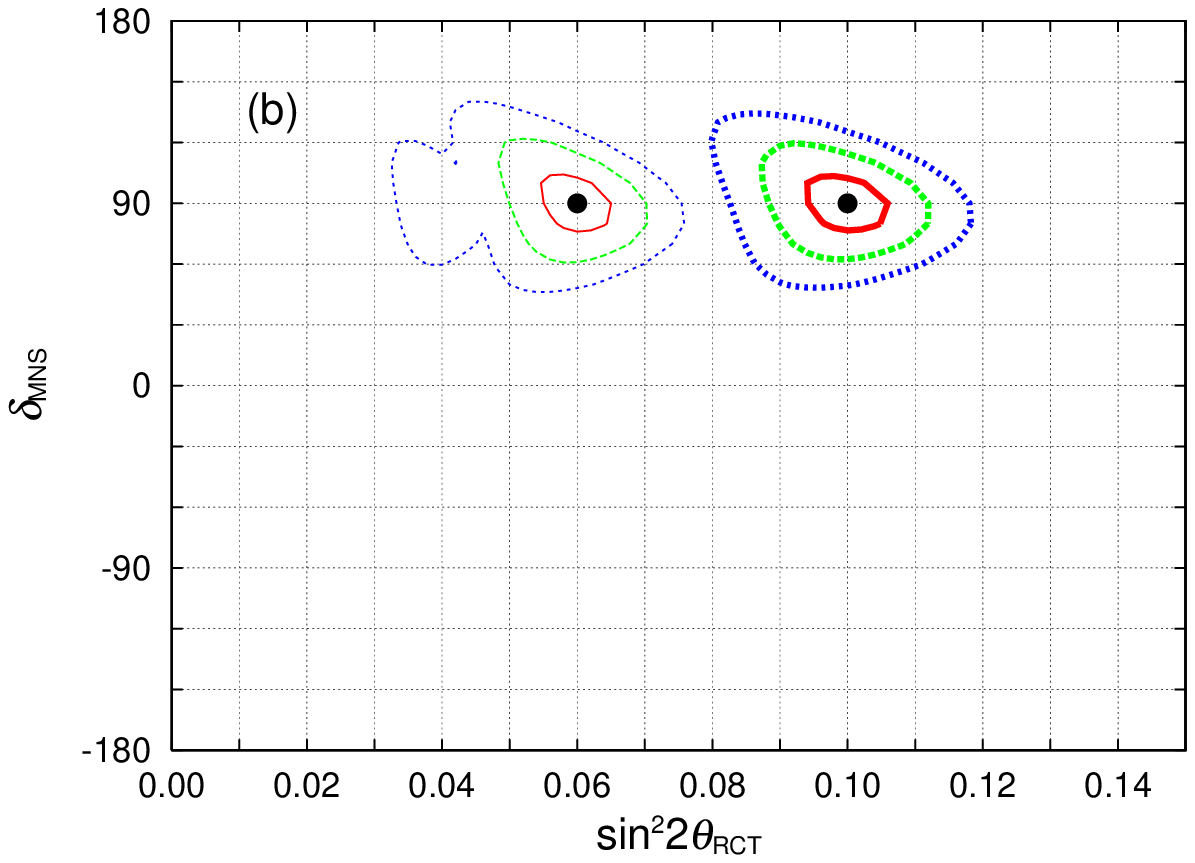}
~~~
  \includegraphics[scale=0.6]{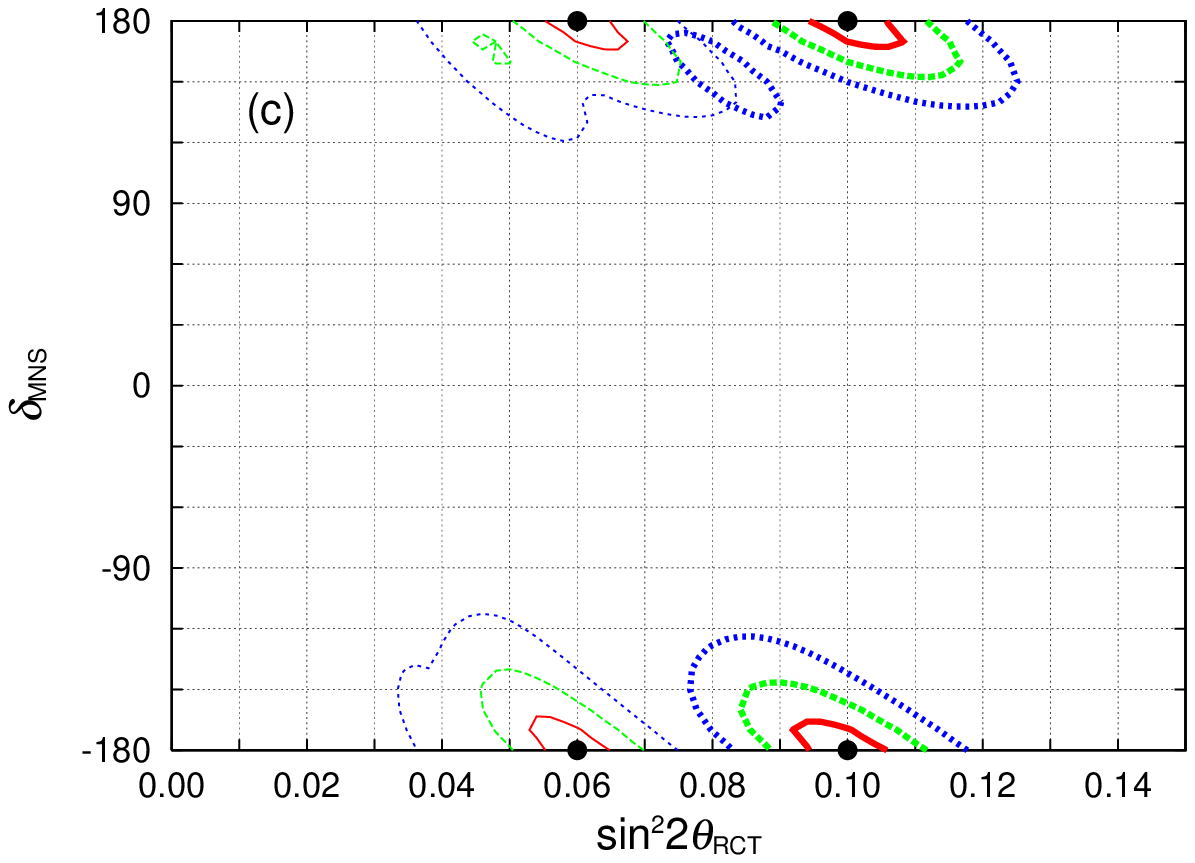}
  \includegraphics[scale=0.6]{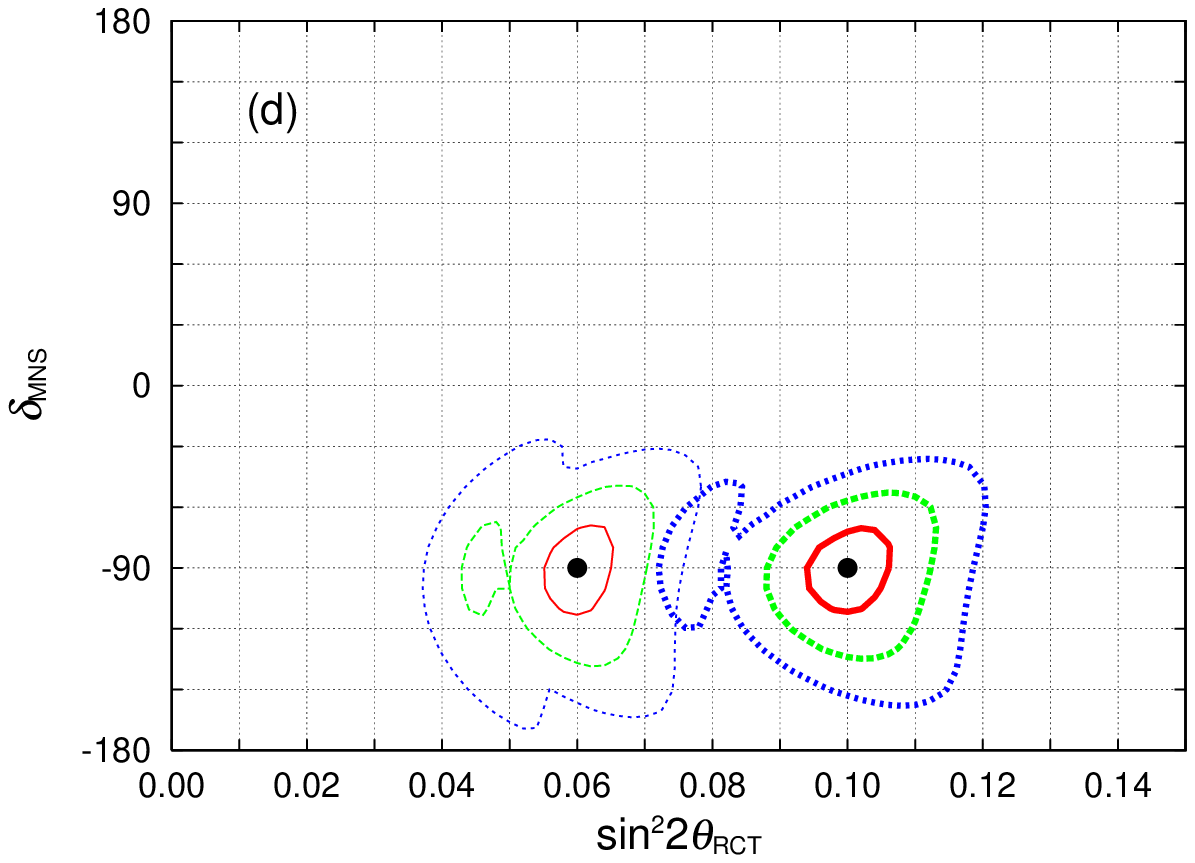}
 \caption{{%
 The allowed region in the plane of
$\srct{2}$ and $\dmns$ by the T2KK set up of \Fgref{oct.chi.wirct},
when $\satm{}^{\rm input}=0.40$ and the hierarchy is normal.
The input values of $\srct{2}$  and $\dmns$,
$\srct{2}^{\rm input}=0.06$ or 0.10,
$\dmns^{\rm input}=0^{\circ}$ (a),
$90^{\circ}$ (b),
$180^{\circ}$ (c), and
$-90^{\circ}$ (d),
are denoted
by solid blobs in each figure 
and the other input parameters are listed in \eqref{input_set}.
The $1\sigma$, $2\sigma$, and $3\sigma$ contours are shown by
solid, dashed, and dotted lines, respectively.
The thick (thin) lines are for $\srct{2}^{\rm input}=0.10$ (0.06). 
There is no allowed region within $3\sigma$
when the inverted hierarchy is assumed with fit.
}}
\Fglab{CP.wirct.040}
\end{figure}
\Figref{CP.wirct.040} shows the potential
of the T2KK experiment for measuring  
$\srct{2}$ and $\dmns{}$  
when $\satm{}^{\rm input}=0.40$ and the hierarchy is normal,
$m_3^2-m_1^2>0$.
The input values of $\srct{2}$ and $\dmns$ are denoted by
solid blobs in each figure,
and the
$1\sigma$, $2\sigma$, and $3\sigma$ allowed regions are
shown by solid, dashed, and dotted lines, respectively.
The thick contours are for $\srct{2}^{\rm input}=0.10$,
and 
the thin contours are for $\srct{2}^{\rm input}=0.06$.
$\dmns^{\rm input}=0^\circ$ in \Fgref{CP.wirct.040}(a),
$90^\circ$ in (b),
$180^\circ$ in (c),
and
$-90^\circ$ in (d).
There is no additional allowed region within $3\sigma$
when the inverted hierarchy is assumed in the fit,
in accordance with \Fgref{chi.cont.mass}(a1).

When we compare the contours of \Fgref{CP.wirct.040} with 
the corresponding ones in Fig.8 of Ref.~\cite{HOS2}
for $\satm{}^{\rm input}=0.5$,
we can clearly identify the islands due to the octant degeneracy.
When $\srct{2}^{\rm input}=0.10$, the thick contours shows no island
at $\dmns^{\rm input}=90^\circ$, but $3\sigma$ islands appear at
all the other $\dmns^{\rm input}$ cases,
which is consistent with the result of \Fgref{oct.cont}(a).
In case of $\srct{}^{\rm input}=0.06$, the contours have $3\sigma$
islands for all $\dmns^{\rm input}$ cases and a clear $2\sigma$ island
at $\dmns^{\rm input}=-90^\circ$, again consistent with
\Fgref{oct.cont}(a).
Close examination of the location of the islands reveals that
their center is at around 
$\srct{2}^{\rm fit} = \srct{2}^{\rm input}~(0.4)/(0.6)$,
as expected by \eqref{mirror}.
The existence of the islands due to the octant degeneracy hence
reduces our capability of measuring $\srct{2}$ significantly.

It is remarkable that the octant degeneracy does not
jeopardize the T2KK capability of determining the CP phase, $\dmns$:
the islands in \Fgref{CP.wirct.040} have the $\dmns$ values
consistent with its input values.
This is because the coefficients of both $\sin \dmns$ and
$\cos \dmns$ in the $\nu_\mu \to \nu_e$ oscillation probability
is not sensitive to the octant degeneracy, as explained in \eqref{q-CP}.
The results we found in \Fgref{CP.wirct.040} confirms the validity
of our approximation for the T2KK experiments.

\begin{figure}[t]
 \centering
  \includegraphics[scale=0.6]{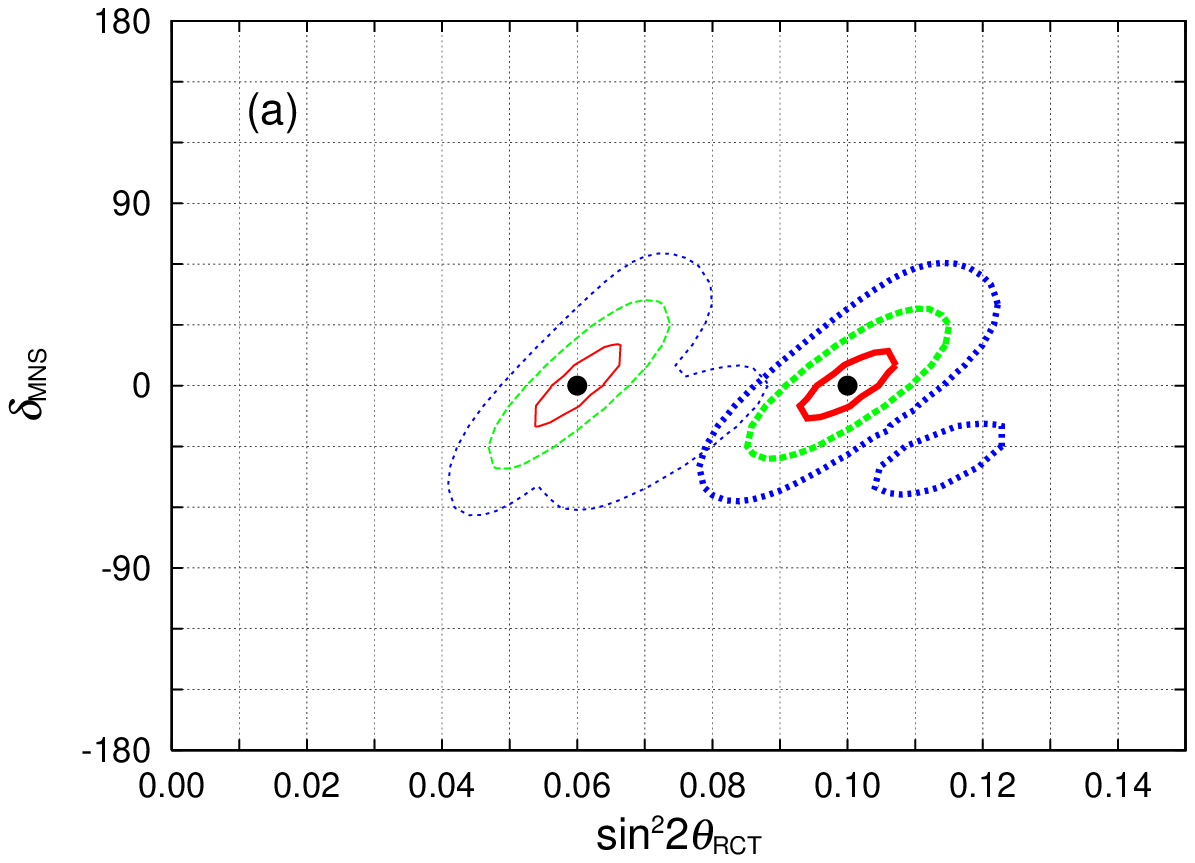}
  \includegraphics[scale=0.6]{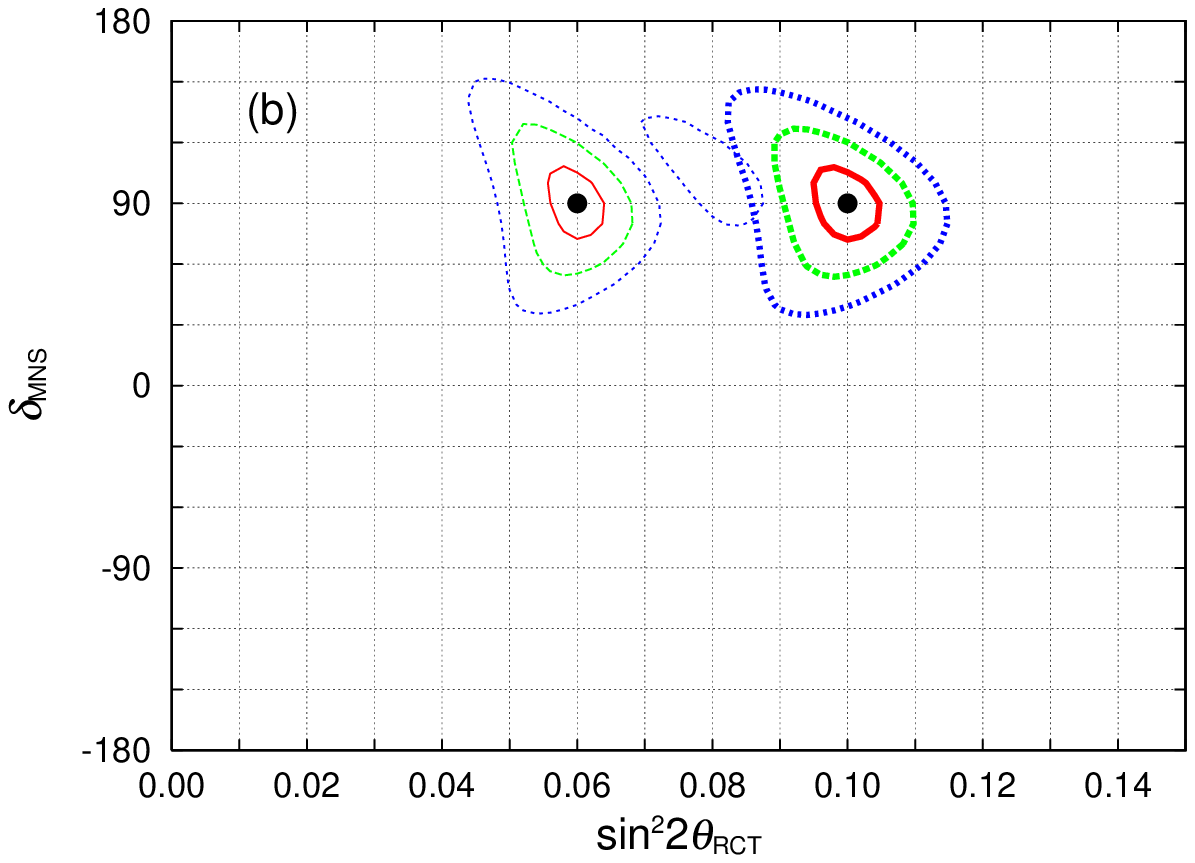}
~~~
  \includegraphics[scale=0.6]{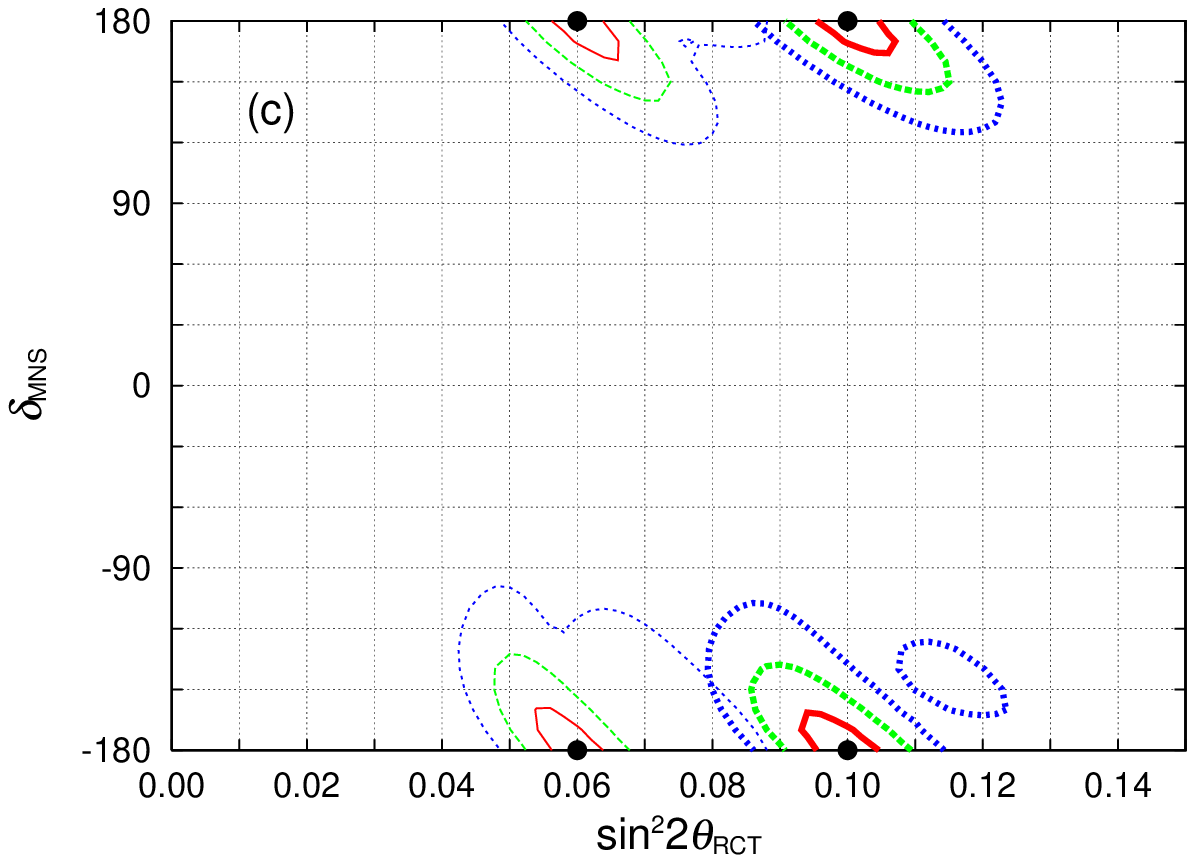}
  \includegraphics[scale=0.6]{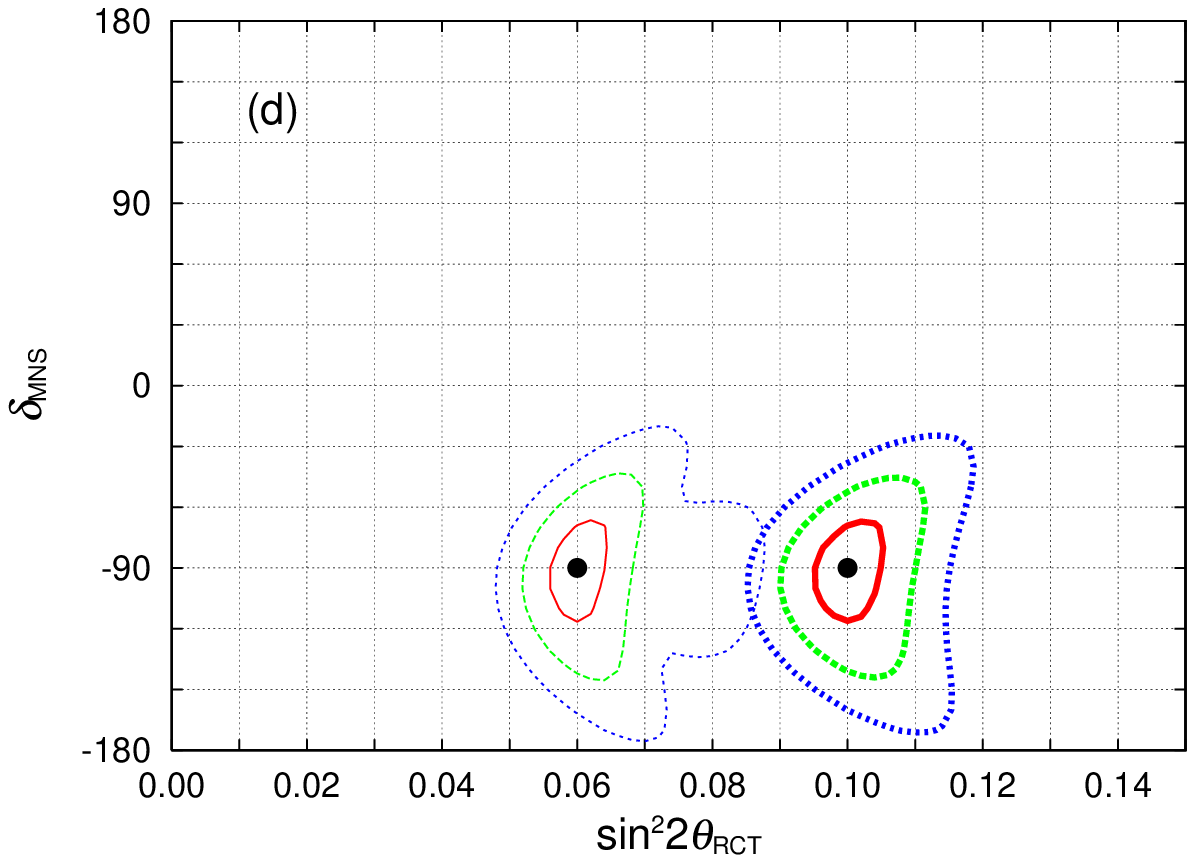}
 \caption{{%
The same as \Fgref{CP.wirct.040}, but
for $\satm{}^{\rm input}=0.60$.
}}
\Fglab{CP.wirct.060}
\end{figure}
\Figref{CP.wirct.060} also shows the potential
of the T2KK experiment for measuring  
$\srct{2}$ and $\dmns{}$, 
but for $\satm{}^{\rm input}=0.60$.
There is no allowed region within $3\sigma$
when the inverted hierarchy is assumed in the fit,
as can be seen from \Fgref{chi.cont.mass}(a3).

Comparing the contours of \Fgref{CP.wirct.060} with 
the corresponding ones in Fig.8 of Ref.~\cite{HOS2}
for $\satm{}^{\rm input}=0.5$,
we can again identify the islands due to the octant degeneracy.
When $\srct{2}^{\rm input}=0.10$, the thick contours shows no island
at $\dmns^{\rm input}=\pm90^\circ$, but $3\sigma$ islands appear at
the other $\dmns^{\rm input}$ cases.
In case of $\srct{}^{\rm input}=0.06$, the contours have $3\sigma$
islands for all $\dmns^{\rm input}$ cases, 
but there is no $2\sigma$ island.
These results are consistent with the result of \Fgref{oct.cont}(a).
The location of the center of the islands is at around
$\srct{2}^{\rm fit} = \srct{2}^{\rm input}~(0.6)/(0.4)$,
as expected by \eqref{mirror}.
The capability of measuring $\srct{2}$ for $\satm{}^{\rm input}=0.60$ 
is also reduced by the octant degeneracy, which is the same as
that for $\satm{}^{\rm input}=0.40$.
However, the octant degeneracy does not disturb
the T2KK capability of determining the CP phase.

\section{Summary}
\label{sec:last}
There are three types of ambiguity in the neutrino parameter space.
The first one is from the sign of the larger mass-squared difference,
which is related to the mass hierarchy pattern.
The second one is from the combination of the unmeasured parameters,
the leptonic CP phase ($\dmns$) and 
the mixing angle $\theta_{\rct}$,
which resides at the upper-right corner of the MNS matrix \cite{MNS}.
In the previous studies \cite{HOS1,HOS2},
we showed that the idea \cite{HOSetc} of placing
a 100kton-level water \cerenkov detector in Korea
along the T2K neutrino beam at $L=1000$km,
the T2KK experiment \cite{korean} can solve
these ambiguities.
The last ambiguity is in the value of the $\theta_{\atm}$,
which dictates the atmospheric neutrino observation \cite{atm}
and the long base-line neutrino oscillation experiment
\cite{k2k,minos}.
If the mixing angle $\theta_{\atm}$, is not $45^\circ$,
there is a two fold ambiguity between
``$\theta_{\atm}$'' and ``$90^\circ-\theta_{\atm}$'',
the octant degeneracy \cite{octant}.

In this paper,
we focus on the physics potential of the T2KK experiment
for solving the octant degeneracy.
In our semi-quantitatively analysis, we follow the strategy of 
Ref.~\cite{HOS1,HOS2} where we adopt SK as a near side detector
and postulate a 100 kton water \cerenkov detector at $L=1000$km,
and the J-PARC neutrino beam orientation is adjusted to 
$3.0^\circ$ at SK and $0.5^\circ$ at the Korean detector site.

 If the value of $\satm{2}$ is 0.99, which is 1\% smaller than the
maximal mixing,
 the value of $\satm{}$ is $\satm{}=0.45$ or $0.55$,
which differ by 20\%.
 Therefore,
we also investigate the impacts of the octant degeneracy
on the physics potential for 
the mass hierarchy determination
and the CP phase measurement by T2KK,
because the leading term of $\nu_\mu \to \nu_e$ oscillation
probability is proportional to $\satm{}$, not $\satm{2}$.

 When we include the constraint for the value of $\srct{}$,
which will be obtained from the future reactor experiments
\cite{KASKA}.
 the octant degeneracy between $\satm{}=$
0.40 and 0.60 can be resolved
at $3\sigma$ level for $\srct{2}>0.12$ (0.08)
after 5 years exposure with $\numt{1.0~(5.0)}{21}$ POT/year,
if the hierarchy is normal,
see \Fgref{oct.chi.wirct} and \Fgref{oct.cont}.
 We find that the contribution from the second maximum of
the $\nu_e^{}\to\nu_\mu^{}$ oscillation probability
at the far detector ($L=1000$km)
plays an important role for solving the octant degeneracy.
 It is also found that 
the octant degeneracy cannot be solved without
the contribution from future reactor experiments.

We also investigate the impact of the octant degeneracy
in the determination of the mass hierarchy pattern.
The T2KK power of resolving the mass hierarchy pattern is
proportional to the value of $\satm{}^{\rm input}$
for the normal hierarchy, see \Fgref{chi.mass.nor},
because the $\nu_\mu \to \nu_e$ rate is proportional to
$\satm{}^{\rm input}$.
When the mass hierarchy is normal, 
we can determine
the mass hierarchy at $3\sigma$ level 
for $\srct{2}^{\rm input} \gsim 0.07$ if $\satm{}^{\rm input}=0.40$,
\Fgref{chi.cont.mass}(a1),
whereas the same holds
for $\srct{2}^{\rm input} \gsim 0.04$ if $\satm{}^{\rm input}=0.60$,
\Fgref{chi.cont.mass}(a3).
On the other hand, 
if the hierarchy is inverted,
$\satm{}=0.5$ is found to be the optimal case for
the mass hierarchy determination,
see \Fgref{chi.mass.inv},
because of
the lower rate of the
$\nu_\mu \to \nu_e$ events for $\satm{}^{\rm input} <0.5$ 
and
the octant degeneracy for $\satm{}^{\rm input} >0.5$.

Finally, we check the effect of the octant degeneracy for
the CP phase measurement, see 
\Fgref{CP.wirct.040}
and
\Fgref{CP.wirct.060}.
The CP phase can be constrained to $\pm 30^\circ$ at $1\sigma$ 
level for $\satm{2}=0.96$, even if we cannot distinguish
between $\satm{}=0.4$ or 0.6.
The error does not increase from that for $\satm{}=0.5$,
because the coefficients of both sine and cosine term of $\dmns{}$
in the $\nu_\mu \to \nu_e$ oscillation probability 
are not sensitive to the octant degeneracy.

\bs

{\it Acknowledgments}\\
We thank our colleagues Y.~Hayato,
A.K.~Ichikawa, K.~Kaneyuki, T.~Kobayashi, and T.~Nakaya,
from whom we learn about K2K and T2K experiments.
We are also grateful to K-i.~Senda and T.~Takeuchi 
for useful discussions and comments.
The work is supported in part by the Core University Program of JSPS.
The numerical calculations were carried out on Altix3700 BX2 at
YITP in Kyoto University.

\bs
{\it Note Added}\\
When finalizing the manuscript,
we learned that a similar study has been performed by
T.~Kajita, {\etal} \cite{kajita}.

\end{document}